\documentstyle[12pt]{article}
\input psfig

\def\xslide#1#2#3#4#5#6#7{\centerline{\psfig
{figure=#1,height=#2,bbllx=#3bp,bblly=#4bp,bburx=#5bp,bbury=#6bp,width=#7,clip=}}}

\def\l{{\bf l}}

\newcommand{\tjnaf}{TJNAF}
\newcommand{\mm}{$m\,\overline{m}$ }
\newcommand{\kk}{$K\overline{K}$ }
\newcommand{\kpkm}{$K^+K^-$ }
\newcommand{\pp}{$\pi^+\pi^-$ }
\newcommand{\pnpn}{$\pi^0\pi^0$ }
\newcommand{\reactionpol}{$\pi^- p_{\uparrow} \rightarrow \pi^+ \pi^- n$ }
\newcommand{\be}{\begin{equation}}
\newcommand{\ee}{\end{equation}}

\begin{document}

\title{Coupled Channel Analysis of 
S-Wave ${\pi}{\pi}$ and \kk Photoproduction} 
\author{Chueng-Ryong Ji$^{a}$, Robert Kami{\'n}ski$^{b}$, 
Leonard Le{\'s}niak$^{b}$,\\
Adam Szczepaniak$^{c}$ and Robert Williams$^{d}$\\ \\
{\em $^{a}$Department of  Physics, Box 8202,
North Carolina}\\ {\em State University,
Raleigh, North Carolina 27695, USA}\\
{\em $^{b}$Department of Theoretical Physics,
H. Niewodnicza{\'n}ski}\\ {\em Institute of Nuclear Physics,
PL 31-342 Krak{\'o}w, Poland}\\
{\em $^{c}$Physics Department,
Indiana University,}\\ {\em Bloomington, Indiana 47405, USA}\\
{\em $^{d}$\tjnaf, 12000 Jefferson Avenue, 
Newport News,}\\ {\em Virginia 23606, USA}}

\maketitle 
\begin{abstract}
We present a coupled channel
partial wave analysis of nondiffractive 
$S-$wave $\pi^+ \pi^-$ and $K^+ K^-$
photoproduction focusing on the 
 \kk threshold.
 Final state interactions are included. 
We calculate total cross sections, angular and effective mass 
 distributions in both $\pi\pi$ and \kk channels. 
Our results indicate that these processes are experimentally
measurable and valuable information on the $f_0$(980) resonance
structure can be obtained.
\end{abstract}

\section{Introduction}
Photoproduction of $K^+ K^-$ pairs near threshold was previously measured
at Daresbury 
\cite{barberetal} and Hamburg \cite{Friesetal}.
An interesting result from these experiments was an observation
of an asymmetry in the $K^+ K^-$ angular distribution 
 in the invariant mass region close to the threshold. This could be 
described as an interference of the dominant $P$-wave from the decay 
 of the $\phi(1020)$ meson and an $S$-wave from the decay of the $f_0(980)$ 
resonance.  
 Furthermore, the data \cite{emaa,behrend,aston} agree 
 with the diffractive $\phi$ meson  
 production mechanism. The $f_{0}$ meson is most likely 
 produced nondiffractively since it has different quantum numbers from the
 photon. 

The scalar meson sector of the hadron spectroscopy is still 
very poorly known.
There exist a variety of  theoretical models dealing with the structure 
of scalar mesons. Even in the ordinary quark model a unique
$q\overline{q}$ assignment of the scalar $^3P_0$ meson nonet seems to be 
impossible \cite{pdg92}. This is expected since glueballs can mix with 
 the scalars  \cite{close,amsovich} and naturally lead to a complicated 
 internal structure. 
Among the scalar mesons  
 isoscalar $f_{0}(980)$ and its isovector partner $a_{0}(980)$ 
are particularly interesting \cite{panic,ichep}.
 The $f_{0}(980)$  branching ratio to $K\overline{K}$ 
  amounts to 22$\%$ which is astonishingly large considering that its mass 
 is below the \kk threshold as most experimental data suggest. 
Therefore theoretical models have been formulated in which the $f_{0}(980)$ 
 is a quasibound state of the $K\overline{K}$ pair \cite{weinstein}.

One may assume that the $K\overline{K}$ pair can be bound in a similar way 
as the nuclear forces bind the 
simplest nucleus -- deuteron. There is, however, 
an important difference between
the deuteron and the $K\overline{K}$ quasibound state, 
namely the $K\overline{K}$ quasibound state can 
annihilate into a system of two pions ($\pi^+\pi^-$ or $\pi^0\pi^0$).
Models for a coupled channel pion--pion and kaon--kaon  scattering 
 like the one in Ref.\cite{klm},  
 have allowed us to obtain some 
characteristics of the $f_{0}(980)$ meson such as  its binding energy,  
 decay width, root 
mean square radius,  branching ratios and various threshold parameters 
\cite{kl}. 
Calculations  of the  scalar meson properties  have been based, however, 
 on the experimental data which are  highly imprecise and even controversial,
particularly for the $K\overline{K}$ production near threshold
\cite{klm}. This is very unsatisfactory if one attempts to fix the 
theoretical parameters by a comparison with experiment. This situation may be
improved in the near future with new data on $f_{0}(980)$ 
photo- and electroproduction experiments 
 from \tjnaf \cite{proposal}. This goal can be accomplished  
if a very good effective mass resolution near the $K\overline{K}$ 
threshold is achieved and a partial wave analysis is
performed in different final state channels.

With this background, it is clear that much more 
extensive and systematic investigations
must be performed on the theoretical side.
Thus, in this paper, we present a coupled channel
$S$-wave analysis of nondiffractive $\pi^+ \pi^-$ and $K^+ K^-$
photoproduction.
The internal structure of the scalar meson $f_{0}(980)$ can be studied 
via the final state interactions of the produced pion or kaon pairs. 
 We expect channel coupling effects to be very important since
the final state interaction (FSI) of two pions (kaons) 
can lead to a production of two kaons (pions) 
in the resonant $S$--wave\cite{klm}. 
A similar dependence on channel coupling can be seen in $a_{0}(980)$ 
production leading to
a formation of the kaon pairs from the initial $\eta$--$\pi$ system 
\cite{atkinson}, however, the isovector channel will not be discussed in 
this paper. 

At low photon energies the $S$--channel 
photon--nucleon resonances are expected to give 
 a substantial contribution 
to  the production amplitude. In Ref.\cite{to} G{\'o}mez Tejedor 
and Oset made a comprehensive analysis of 
 $\gamma p \rightarrow \pi^+ \pi^- p$  including 
 contributions from  $\Delta(1232), N^* (1440)$ and $N^* (1520)$ resonances. 
The model reproduced experimental cross sections fairly well below 
$E_\gamma = 800$  MeV
and invariant-mass distributions at even slightly higher energies. 
 In the analysis of forward $\pi^+ \pi^-$ and 
$K^+ K^-$ photoproduction at much higher energies 
($E_{\gamma} \stackrel{>}{\sim} 4 \;$ GeV) 
we expect however that mostly the $t$--channel $\rho$ and $\omega$ meson
exchanges become  important. 
 This assumption follows from
Regge phenomenology where high energy amplitudes are dominated by
$t$-channel meson exchange. 
In this paper, we focus on the coupled channel $S$-wave
analysis of forward $\pi^+ \pi^-$ and $K^+ K^-$ photoproduction
near the $K {\overline K}$ threshold and 
thus we limit the calculation of the nonresonant two meson production 
  amplitudes to a  few 
  Feynman diagrams, schematically shown in Fig.~1.
 
The paper is organized as follows. 
In the next section we calculate the amplitude corresponding to the Feynman
diagrams of Fig.~1 and show how they are related to the differential
cross section in the Born approximation. We also give the  partial $S$-wave
decomposition of the invariant amplitudes for
$\pi^+ \pi^-, \pi^0 \pi^0$ and $K^+ K^-$ production.
In section~III we make an off--shell extension of the Born amplitudes and 
relate them to the photoproduction potential. 
The full amplitudes including 
the channel coupling and final state interactions are also discussed in 
 section~III.
In section~IV  numerical results are
presented and parameter dependence is addressed. 
 The conclusions and discussions follow in section~V.

In the Appendices we summarize the effective interaction Lagrangian
used in this analysis (Appendix A), the detailed factors in the current defined in 
section II (Appendix B) and the results of the $S$-wave amplitude
(Appendix C).

\section{Nonresonant Two Meson  Production Amplitude and Differential Cross 
Section}

As discussed in the introduction, we  focus on the coupled channel 
$S$-wave analysis of nondiffractive $\pi^+ \pi^-$ and $K^+ K^-$ photoproduction
near and above the $K {\overline K}$ threshold of the ${f_0}(980)$ meson.
According to the Regge phenomenology,
we expect that only the $t$-channel meson exchanges, 
especially $\rho$ and $\omega$ mesons,
are important in the analysis of the $S$-wave forward $\pi^+ \pi^-$ and 
$K^+ K^-$ photoproduction at higher energies.
However, the pion coupling to proton cannot contribute because our 
calculations are limited only to the $S$-wave final states.
As illustrated in Fig.~1, 
we calculate five diagrams for the 
$\pi^+ \pi^-$ production and six diagrams for the 
$K^+ K^-$ production. 
Among the five  diagrams for the $\pi^+ \pi^-$ production, three diagrams
correspond to $a = \pi^{\pm}$ and $b = \rho^0$  
including a contact diagram as required by current conservation.  
In the remaining two diagrams $a = \rho^{\pm}$ and $b = \omega$. 
Later, we will refer to the first three diagrams, including the contact term, 
 as the type $I$ and to these with double vector meson exchanges in the t-channel 
 as the type $II$. 
 For the six diagrams of
$K^+ K^-$ production,  three of them correspond to $a = K^{\pm}$ and 
 $b = \rho^0$
(type $I$) and the other three to $a = K^{\pm}$ and $b = \omega$ 
(again type $I$). 

Using the effective interaction Lagrangians listed
in the Appendix A, we first define the on-shell meson pair 
($\pi \pi$ or $K {\bar K}$)
production amplitude through the appropriate Born terms. 
Similar effective hadronic field theories have been extensively used in the
single kaon electro- and photoproduction processes \cite{crji}. 
For example, for the double pion photoproduction
{\it i.e.} $\gamma p \to \pi\pi p$, we have 
\begin{eqnarray}
&&-i(2\pi)^4\delta(p'+k_1+k_2-p-q)V_{\pi^i\pi^j}(k_1,k_2,p',p,q,s',s,\lambda)
\nonumber \\
&&= 
\langle \pi^i(k_1)\pi^j(k_2)\{p'\}|Te^{i\int dx {\cal L}}|\{p\},\{\gamma\}
\rangle,
\label{evol}
\end{eqnarray}
where $i,j=\pm,0$ refer to pion charge, 
 $\{\gamma\}=(q,\lambda)$ is the set of quantum numbers representing the 
incoming photon  momentum and spin projection. 
The initial and final proton states are parameterized by 
  $\{p\}=(p,s)$ and $\{p'\} =(p',s')$, respectively.
For $\gamma p \to \pi^+ \pi^- p$  the 
expansion of the evolution operator with the Lagrangian ${\cal L}$ given 
in the Appendix A yields five diagrams
with the $t$-channel exchange as mentioned above, {\it i.e.}
three diagrams with $a = \pi^{\pm}$ and $b = \rho^0$ (type $I$)
and two diagrams
with $a = \rho^{\pm}$ and $b = \omega$ (type $II$) in Fig.~1.
Similarly, by expanding the evolution operator in Eq.(\ref{evol})
for $\gamma p \to \pi^0 \pi^0 p$  we only obtain two type-$II$ processes with 
  $(a,b)=(\rho^0,\omega)$
and $(a,b)=(\omega,\rho^0)$. 
For the $K^+ K^-$ photoproduction, the expansion of the
evolution operator in Eq.(\ref{evol}) generates six diagrams,
all of which belong to the type $I$
as we have discussed earlier,{\it i.e.} three diagrams with
$(a,b)=(K^{\pm},\rho^0)$ plus three diagrams with $(a,b)=(K^{\pm},\omega)$.

Now, using the notations $m$ and ${\bar m}$ for the two produced mesons 
 the total amplitude can be written 
as 
\begin{equation}
V_{m\bar m} = \sum_{r= I,II} \bar u(p',s')J_{r,m \bar m}\cdot 
\epsilon(q,\lambda)u(p,s),
\label{Vmm}
\end{equation}
where $r = I$ refers to sum over the three diagrams
including the contact diagram and $r = II$
denotes the sum over
only the two diagrams without the contact diagram as
discussed earlier.
There are three cases of meson pair photoproduction in this analysis,
\mm $= {\pi^+ \pi^-}, {\pi^0 \pi^0}, {K^+ K^-}$.
The summary of calculated diagrams for each case of \mm   is listed
in Table 1. 
 The main part of $V_{m\bar m}$ is of course  the hadronic current 
$J_{r,m\bar m}^{\mu}$ multiplied by the photon polarization four--vector 
$\varepsilon(q,\lambda)$. The initial 
and final proton spinors are denoted by $u(p,s)$ and $\bar u(p',s')$.
Calculated expressions for $J_{r,m\bar m}^{\mu}$ are given by
\begin{equation}
J_{r,m\bar m}^{\mu} = (\alpha_{r,m\bar m} g^{\mu \nu} + k_1^\mu 
\beta_{1r,m\bar m}^\nu
+ k_2^\mu \beta_{2r,m\bar m}^\nu)\{d_{r,m\bar m} \gamma_\nu + e_{r,m \bar m}
 (p+p')_\nu\},
\label{J}
\end{equation}
where the variables 
$\alpha_{r,m\bar m}, \beta_{1r,m\bar m}, \beta_{2r,m\bar m}, d_{r,m\bar m}$ 
and $e_{r,m \bar m}$ are written in 
the Appendix B. Here, $g^{\mu\nu}$ is the metric tensor 
($g^{00} =1, g^{11} = g^{22} = g^{33} = -1$) and $\gamma^{\nu}$ are Dirac matrices.
The 
 Born double differential cross section is then given by 
\begin{equation}
\frac{d\sigma}{dtdM_{m\bar m}}
= {1\over {(2\pi)^3}}{ {|\vec\kappa|} \over {32m_p^2\omega^2}}
\sum_{l,l_z}\langle|V^{l,l_z}_{m\bar m}|^2 \rangle,
\label{dsigma}
\end{equation}
where $V_{m\bar m}^{l,l_z}$ is the partial-wave projected 
amplitude for the process $\gamma p \rightarrow
m \bar m p$, 
\begin{equation}
V^{l,l_z}_{m\bar m} = \int d\Omega_{m\bar m}Y_{ll_z}(\Omega_{m\bar m})
V_{m\bar m}.
\label{partial}
\end{equation}
Here $l$ is the angular momentum of the relative \mm  motion,
$l_z$ is its projection, $t=(p'-p)^2$ is the four--momentum transfer squared 
and $M_{m \bar m} = \sqrt{(k_1+k_2)^2}$ is the two--meson effective mass. 
In Eqs. (\ref{dsigma}) and (\ref{partial}) 
$|\vec\kappa| = \sqrt{M^2_{m\bar m}/4-m_m^2}$ and 
$\Omega_{m\bar m}$ is the angle between the relative momentum of the 
 two mesons in their c.m. frame and the direction of the final proton momentum.
 The photon energy in the laboratory 
frame is denoted by $\omega$ and 
$m_p$ is the proton mass. 

The spin averaged amplitude square in Eq.(\ref{dsigma}) is given by
\begin{equation}
\langle |V^{ll_z}_{m\bar m}|^2 \rangle = \frac{1}{4} \sum_{s,s',\lambda}
 |\langle s'|
V^{ll_z}_{m\bar m}|s,\lambda\rangle|^2.
\label{Tsquare}
\end{equation}
Furthermore, because only  terms
inside the round bracket in Eq.(\ref{J}) depend on meson momenta,
the partial-wave projected Born amplitude can also be written as

\begin{equation}
V^{ll_z}_{m\bar m} = 
\sum_r \bar u(p',s')J^{ll_z}_{r,m\bar m}\cdot \epsilon(q,\lambda)
u(p,s),
\end{equation}
where
\begin{equation}
J^{ll_z,\mu}_{r,m\bar m} = P^{ll_z,\mu\nu}_{r,m\bar m}
 \{d_{r,m\bar m} \gamma_\nu + e_{r,m\bar m} (p+p')_\nu\},
 \label{Jtilde}
\end{equation}
and the tensor $P$ is given by 
\begin{equation}
P^{ll_z,\mu\nu}_{r,m\bar m} = \int d\Omega_{m\bar m}
Y_{llz}^*(\Omega_{m\bar m}) (\alpha_{r,m \bar m} g^{\mu \nu} +
k_1^\mu \beta_{1r,m\bar m}^\nu + k_2^\mu \beta_{2r,m\bar m}^\nu).
\label{stensor}
\end{equation}
The $S$-wave results of the integration over the solid angle 
in Eq.(\ref{stensor}) are
summarized in the Appendix C.

\section{Final State Interactions}
We now discuss  the inclusion of final state interactions (FSI)
to the $S$-wave projected amplitudes $l,l_z=0,0$. Since we 
 are only interested in the 
 $S$-wave projection we shall drop the 
partial wave labels $l,l_z$ from now on. 
We also limit the study to 
  $\pi^+\pi^-,\pi^0\pi^0$ and $K^+ K^-$ coupled channel interactions
in the final state.

The full photoproduction amplitude including final state interactions can 
be written in the operator form as 

\begin{equation}
{\hat T} = {\hat V} + 
{\hat t} \hat G {\hat V}, \label {FFSI}
\end{equation}
where ${\hat V}$ is the photoproduction potential, $\hat t$ is the strong 
FSI t-matrix, and $\hat G$ is the propagator of the intermediate state. 
The matrix elements of $\hat V$ are obtained through an off-shell extension 
of the Born amplitude. This is done by following Ref. \cite{Bakker}  
and generalizing the relativistic 
scattering problem to inelastic channels. 
 The final result is the potential $\hat V$ obtained as  
 the off-energy-shell extension of the 
Born amplitudes evaluated in the total c.m. frame. This photoproduction 
potential is then coupled to FSI t-matrix in the two meson rest frame. 
For the $\gamma p\to m \bar m p $ photoproduction the matrix element 
of Eq.~(\ref{FFSI}) is given by

\begin{eqnarray}
T_{m\bar m} &=& 
\langle m\bar m,\vec\kappa,\{p'\}|\hat T|\{p\},\{q\}\rangle 
\nonumber \\
&=& 
\langle m\bar m,\vec\kappa,\{p'\}|\hat V|\{p\},\{q\}\rangle
\nonumber \\
&+& 
4\pi \sum_{m'\bar m'}\int_0^\infty {{\kappa'^2d\kappa'}\over {(2\pi)^3}}  
F(\kappa,\kappa') 
\langle m\bar m,\vec\kappa|\hat t| m' \bar m', 
\vec\kappa' \rangle G_{m'\bar m'}(\vec\kappa')
\nonumber \\
&&<m'\bar m',\vec\kappa',\{p'\}|\hat V|\{p\},\{\gamma\}>,
\label{fs}
\end{eqnarray}
where
\begin{equation}
F(\kappa,\kappa') = \sqrt{\frac{M_{m\bar m}E_{m\bar m}}{M'_{m'\bar m'}
E'_{m'\bar m'}}},
\end{equation}

\begin{equation}
M'_{m'\bar m'} = 2 \sqrt{m_{m'}^2 + \kappa'^2}, \;\;
E_{m\bar m} = \sqrt{M_{m\bar m}^2 + p^2}, \;\;
E'_{m'\bar m'} = \sqrt{{M'}^2_{m'\bar m'} + {p}^2},
\end{equation}
and 
\begin{equation}
G_{m'\bar m'}(\kappa') = 
{1 \over {M_{m\bar m}-M'_{m'\bar m'}(\kappa')+i\epsilon}} \; .
\label{prop}
\end{equation}
Here, $p$ is the two meson  momentum in the overall c.m. 
system. 
The half-off-shell behavior of the photoproduction potential will in 
general be affected by strong vertex form factors. We take this effect 
into account by employing a global form factor so that 
\begin{equation}
\langle m'\bar m',\vec\kappa',\{p'\} 
|\hat{V}\longrightarrow F_{cut}(M'_{m'\bar m'},M_{m\bar m})
\langle m'\bar m',\vec\kappa',\{p'\} |\hat{V},
\label{vhat}
\end{equation}
where $M'_{m'\bar m'}$
is the invariant 
mass of the two off-shell interme\-dia\-te mesons, and   
 $M_{m\bar m} = 2\sqrt{\kappa^2 + m_{m}^2}$ is the on-shell 
mass of the produced (detected) meson pair.  
The various choices of 
functional forms
for $F_{cut}(M',M)$ sa\-tis\-fying the normalization condition, 
$F_{cut}(M,M) = 1$,  will be discussed
in detail in the next section.

As shown in references \cite{klm,kl}, there are in general more than
one channel that contribute to the production of a given final meson pair.
Thus, the final state interaction leads to a coupled channel problem
for  $T_{m\bar m}$ in Eq.(\ref{fs}). 
The isospin decomposition of each final state requires the inclusion of 
other possible meson pairs such as $\pi^0 \pi^0$ and $K^0 {\bar K}^0$
and thus one has to consider all four channels ($\pi^0 \pi^0$,
$\pi^+ \pi^-$, $K^+ K^-$ and $K^0 {\bar K}^0$) 
as the intermediate
states. 
The $\pi^0 \pi^0$,  
$\pi^+ \pi^-$, $K^+ K^-$ and $K^0 {\bar K}^0$ Born photoproduction
amplitudes are specified as $V_{\pi^0\pi^0},V_{\pi^+\pi^-},V_{K^+ K^-}$ and 
$V_{K^0 {\bar K}^0}$, 
respectively. 
We have classified the final state scattering amplitudes
as the components of a two-by-two matrix in the basis of pion 
and kaon pair states. 
For example, the isospin 0 ($I=0$) $S$-wave two pion state in the meson 
c.m. frame 
(${\vec P}_{\pi\pi}=0$)  is given by 

\begin{equation}
|I=0,S;\pi\pi\rangle = {1\over \sqrt{2}}\int {{d{\vec \kappa}}\over {(2\pi)^3}}
{1\over \sqrt{3}} (|\pi^+({\vec \kappa})\pi^-(-{\vec \kappa})\rangle 
 + |\pi^-({\vec \kappa})\pi^+(-{\vec \kappa})\rangle 
- |\pi^0({\vec k})\pi^0(-{\vec k})\rangle),
\end{equation}
where we have used the convention that under isospin $|\pi^+\rangle$ 
transforms as \\ $+|I=1,I_3=1\rangle$. A similar expression may be obtained 
  for the $I=2$ $S$-wave.
The \kk, $I=0$ $S$-wave state is given by 

\begin{equation}
|I=0,S;K\bar K\rangle = \int {{d{\vec k}}\over {(2\pi)^3}}
{1\over \sqrt{2}} (|K^+({\vec k})K^-(-{\vec k})\rangle 
 -|K^0({\vec k})\bar K^0(-{\vec k})\rangle),
\end{equation}
where the convention is that $(K^+,K^0)$ and $(K^-,\bar K^0)$ form 
isospin doublets. 
The $S$-wave amplitudes $T_{\pi^+\pi^-}$ and 
$T_{\pi^0\pi^0}$ in terms of isospin FSI t-matrix and 
$S$-wave potentials $V_{m\bar m}$ are thus given by

\begin{eqnarray}
T_{\pi^+ \pi^-} &=& [1+ir_{\pi}(\frac{2}{3} { t}_{\pi\pi}^{I=0}
+ \frac{1}{3}{ t}_{\pi\pi}^{I=2}) + \frac{2}{3}\hat{ 
P}_{\pi\pi}^{I=0}
+\frac{1}{3}\hat{ P}_{\pi\pi}^{I=2}] \; V_{\pi^+\pi^-} \nonumber \\
&+& \frac{1}{3}[ ir_\pi(-{ t}_{\pi\pi}^{I=0} + { 
t}_{\pi\pi}^{I=2})
-\hat{ P}_{\pi\pi}^{I=0} + \hat{ P}_{\pi\pi}^{I=2}] \; 
V_{\pi^0 \pi^0} \nonumber \\
&+& \frac{1}{\sqrt{6}}( ir_K {t}_{\pi K}^{I=0} +
\hat{P}_{\pi K}^{I=0})(V_{K^+ K^-} - V_{K^0{\bar K}^0})
\label{pi+-decomp}
\end{eqnarray} 
and
\begin{eqnarray}
T_{\pi^0 \pi^0} &=& [1+ir_{\pi}(\frac{1}{3} {t}_{\pi\pi}^{I=0}
+ \frac{2}{3}{t}_{\pi\pi}^{I=2}) + \frac{1}{3}\hat{ 
P}_{\pi\pi}^{I=0}
+\frac{2}{3}\hat{P}_{\pi\pi}^{I=2}] \; V_{\pi^0\pi^0} \nonumber \\
&+& \frac{2}{3}[ ir_{\pi}(-{t}_{\pi\pi}^{I=0} +
{t}_{\pi\pi}^{I=2})
-\hat{P}_{\pi\pi}^{I=0}+\hat{P}_{\pi\pi}^{I=2}]V_{\pi^+ \pi^-} 
\nonumber  \\
&-& \frac{1}{\sqrt{6}}[ ir_K{t}_{\pi K}^{I=0} + 
\hat{P}_{\pi K}^{I=0}] \; (V_{K^+ K^-} - V_{K^0 {\bar K}^0}),
\label{pi00decomp}
\end{eqnarray} 
where ${\hat {P}}$ is the principal value (PV) 
integration part induced from the
intermediate energy propagators (see Eq.(\ref{fs})). 
 Note that $\hat{P}$ is an integral 
operator which acts on the Born amplitudes and implicitly depends on
the corresponding half off--shell strong interaction ${t}$ matrix
elements. The imaginary part of the intermediate state
propagator leads to an integral over  a  $\delta$-function 
 which produces the coefficients $r_{\pi}$ and $r_K$ given by
$r_{\pi} = - {|{\vec {\kappa}}_\pi| M_{\pi \pi}}/{8 \pi}$
($M_{\pi\pi} = 2 \sqrt{m_{\pi}^2 + {\vec \kappa}_{\pi}^2}$)
and $r_K = - {|{\vec {\kappa}}_K| M_{K K}}/{8 \pi}$ 
($M_{K K} = 2 \sqrt{m_K^2 + {\vec \kappa}_K^2}$).
If we consider the $\pi^+ \pi^-$ photoproduction with the
effective mass smaller than the $K {\bar K}$ threshold
($M_{\pi \pi} < 2 m_K$), then there is no on-shell contribution
from the intermediate $K {\bar K}$ state, {\it i.e.} one should
take $r_K = 0$.
The \pp and \kk elastic scattering amplitudes are denoted by $t_{\pi\pi}$
and $t_{K\overline{K}}$ while $t_{\pi K}$ and $t_{K \pi}$  
are the transition amplitudes for the processes 
$K\overline{K} \longrightarrow \pi\pi$ and
$\pi\pi \longrightarrow K\overline{K}$, respectively.
The intermediate operators ${\hat {P}}_{\pi K}$ and
${\hat {P}}_{K \pi}$ include the propagators of $K {\bar K}$
and $\pi \pi$ pairs, respectively.
The pion and kaon propagators are given by Eq.~(\ref{prop}) 
by setting $m'\bar m' = \pi \pi$ and $K \bar K$, respectively, while
$m \bar m = \pi \pi$ would be fixed in calculations of $T_{\pi^+\pi^-}$
and $T_{\pi^0\pi^0}$.
The $\pi \pi$ photoproduction amplitudes projected onto 
 $I =0$ and $2$ can be written in the following way:
\begin{eqnarray}
T_{\pi \pi}^{I=0} &=& (1 + ir_\pi{t}_{\pi \pi}^{I=0} +
\hat{P}_{\pi \pi}^{I=0}) \; V_{\pi \pi}^{I=0} \nonumber \\
&+& (ir_K{t}_{\pi K}^{I=0} + \hat{P}_{\pi K}^{I=0}) \;
V_{K {\bar K}}^{I=0},
\label{tpipi0}
\end{eqnarray}
\begin{eqnarray}
T_{\pi \pi}^{I=2} &=& (1 + ir_\pi{t}_{\pi \pi}^{I=2} +
\hat{P}_{\pi \pi}^{I=2}) \; V_{\pi \pi}^{I=2},
\end{eqnarray}
where
\begin{equation}
V_{\pi \pi}^{I=0} = -\frac{1}{\sqrt{3}}V_{\pi^0 \pi^0} + 
\frac{2}{\sqrt{3}}V_{\pi^+ \pi^-},
\end{equation}
\begin{equation}
V_{\pi \pi}^{I=2} = \sqrt{\frac{2}{3}}(V_{\pi^0 \pi^0} + V_{\pi^+ \pi^-}),
\end{equation}
\begin{equation}
V_{K {\bar K}}^{I=0} = \frac{1}{\sqrt{2}}(V_{K^+ K^-} - 
V_{K^0 {\bar K}^0}).
\end{equation}
Likewise, the $K {\bar K}$ photoproduction amplitudes
with $I = 0$ and $1$ can be expressed as 
\begin{eqnarray}
T_{K {\bar K}}^{I=0} &=& (1 + ir_K{t}_{K {\bar K}}^{I=0} +
\hat{P}_{K {\bar K}}^{I=0}) \; V_{K {\bar K}}^{I=0} \nonumber \\
&+& (ir_{\pi}{t}_{K \pi}^{I=0} + \hat{P}_{K \pi}^{I=0}) \;
V_{\pi \pi}^{I=0},
\end{eqnarray}
\begin{eqnarray}
T_{K {\bar K}}^{I=1} &=& (1 + ir_K{t}_{K {\bar K}}^{I=1} +
\hat{P}_{K {\bar K}}^{I=1}) \; V_{K {\bar K}}^{I=1},
\end{eqnarray}
where
\begin{equation}
V_{K {\bar K}}^{I=1} = \frac{1}{\sqrt{2}}(V_{K^+ K^-} + 
V_{K^0 {\bar K}^0}).
\end{equation}

Similarly, for the $K {\bar K}$ final states, we obtain 
\begin{eqnarray}
T_{K^+ K^-} &=& [1 + ir_K\frac{1}{2}({t}_{KK}^{I=0}
+{t}_{KK}^{I=1}) + \frac{1}{2}(\hat{P}_{KK}^{I=0} +
\hat{P}_{KK}^{I=1})] \; V_{K^+ K^-} \nonumber \\
&+& \frac{1}{2}[ ir_K (-{t}_{KK}^{I=0} + {t}_{KK}^{I=1})
- \hat{P}_{KK}^{I=0} + \hat{P}_{KK}^{I=1} ] \; V_{K^0 {\bar K^0}} 
\nonumber \\
&+& \sqrt{\frac{2}{3}}( ir_{\pi}{t}_{K \pi}^{I=0} +
\hat{P}_{K \pi}^{I=0} ) \; V_{\pi^+ \pi^-}
- \frac{1}{\sqrt{6}}( ir_\pi{t}_{K \pi}^{I=0}  +
\hat{P}_{K \pi}^{I=0}) \; V_{\pi^0 \pi^0} 
\label{K+-decomp}
\end{eqnarray}
and
\begin{eqnarray}
T_{K^0 {\bar K}^0} &=& [1 + ir_K\frac{1}{2}({t}_{KK}^{I=1}
+{t}_{KK}^{I=0}) + \frac{1}{2}(\hat{P}_{KK}^{I=1} +
\hat{P}_{KK}^{I=0})] \; V_{K^0 {\bar K}^0} \nonumber \\
&+& \frac{1}{2}[ ir_K (-{t}_{KK}^{I=0} + {t}_{KK}^{I=1})
- \hat{P}_{KK}^{I=0} + \hat{P}_{KK}^{I=1} ] \; V_{K^+ K^-} 
\nonumber \\
&-& \sqrt{\frac{2}{3}}( ir_{\pi}{t}_{K \pi}^{I=0} +
\hat{P}_{K \pi}^{I=0} ) \; V_{\pi^+ \pi^-}
+ \frac{1}{\sqrt{6}}( ir_\pi{t}_{K \pi}^{I=0}  +
\hat{P}_{K \pi}^{I=0}) \; V_{\pi^0 \pi^0}. 
\label{K00decomp}
\end{eqnarray}
In numerical calculations, we took the  average values of $m_\pi = \frac{1}{2}
(m_{\pi^0} + m_{\pi^+}) = 137.27$ MeV and 
$m_K = \frac{1}{2}(m_{K^0} + m_{K^+}) = 495.69$ MeV.
 {\it i.e.} we did not consider differences in thresholds of charged
and neutral meson pairs. 
In Ref. \cite{ollerosset}, Oller and Oset assumed that the \kk potentials in 
$I=1$ and $I=0$ channels are different by a factor of 3.
However, it is not clear that this difference would persist at the scattering amplitude
level since in both channels the effects from resonances 
$f_0(980)$ and $a_0(980)$ should be very strong near the \kk thresholds.
Therefore, in the present calculations we have assumed 
${t}_{KK}^{I=1} = {t}_{KK}^{I=0}$ and consequently
$\hat{P}_{KK}^{I=1} = \hat{P}_{KK}^{I=0}$. 
This assumption can be removed if $t^{I=1}_{KK}$ is well 
constrained by future experiments.

In our calculations we use the $\pi \pi$ and $K {\bar K}$ strong
interaction amplitudes derived in Ref.~\cite{klm}.
Explicit expressions for the $I=0$ matrix elements can
be found in the Appendix A of Ref.~\cite{klm}. 
Parameterization of the $I=2$ elastic $\pi \pi$ amplitude is given in 
Ref.~\cite{klr}. 
For the $I=0$ \pp amplitude we use parameters obtained
in a recent analysis of the \reactionpol reaction on polarized target
\cite{kll}. They correspond to the two--channel fit for the so--called 
"down--flat" solution of Ref. \cite{klr}.

\section{Numerical Results}

We proceed with a discussion of the model parameters and then continue with 
analysis of numerical results. Since the $S$-wave amplitude is
assumed to be mediated by single or double $\rho, \omega$ exchanges 
in the t-channel, we require 8 hadronic 
($g_{\rho \pi \pi}$, $g_{\rho K K}$, $g_{\omega K K}$,
 $g_{\omega \rho \pi}$, $G_{V}^{\rho}$, $G_{T}^{\rho}$, $G_{V}^{\omega}$,
 $G_{T}^{\omega}$) and 2 electromagnetic ($g_{\rho \pi \gamma}$,
 $g_{\omega \pi \gamma}$) vector meson coupling constants. 
>From the $\rho \rightarrow \pi \pi$ decay width, we find the
coupling constant $g_{\rho \pi \pi} = 6.05$. We use the SU(3) relations
$g_{\rho K {\bar K}} = g_{\omega K {\bar K}} = \frac{1}{2} g_{\rho \pi \pi}$ 
to fix the kaon couplings. 
The same relations lead to a good description of the kaon form factor
in a framework of the vector dominance model.
We find several quoted values for the 
$g_{\omega \rho \pi}$ coupling in the literature~\cite{Renard,Pilkuhn,Rudaz} 
with values
ranging from 10 GeV$^{-1}\stackrel{<}{\sim} |g_{\omega \rho \pi}| 
\stackrel{<}{\sim}
20$ GeV$^{-1}$. 
We use
$g_{\omega \rho \pi} = 14.0$ GeV$^{-1}$, which is close to the value reported
in Ref.~\cite{Bramon}.   
For the $\rho$ meson vector and tensor couplings to nucleon, we adopt two
sets, one corresponding to the Bonn potential~\cite{Machleidt} 
($G_{V}^{\rho} = 2.27$, $G_{T}^{\rho} = 13.85$) further called Bonn 
parameters, and 
the other taken from a phenomenological analysis of $\pi^{+} \pi^{-}$ 
photoproduction~\cite{to} 
($G_{V}^{\rho} = 2.9$, $G_{T}^{\rho} = 18.15$) which we shall refer as 
the Spanish parameters. 
The tensor $\omega N$ 
coupling is known to be very small, hence we take $G_{T}^{\omega} = 0$.
The vector $\omega N$ coupling is also well determined, and we take the
Bonn potential value of $G_{V}^{\omega} = 11.54$. As in the Bonn potential,
we employ a monopole form factor at the $\rho NN$ and $\omega NN$ vertices, 
$F_{bNN}(t) = \lambda^2/[\lambda^2 - t]$ with the scale parameter 
$\lambda = 1.4 \;$ GeV.
Since the Bonn parameters are determined at the normalization point
$t = 0$ using normal $\rho$ and $\omega$ propagators, 
 we renormalize the corresponding  Regge $\rho NN$ and $\omega NN$
couplings such that they equal the Bonn
couplings at $t = 0$ using the prescription:
\begin{equation}
G_{V,T}^{Regge} = \frac{G_{V,T}}{m_v^2 | \Pi_v(s_0, t=0) |_{Regge}}.
\label{G_regge}
\end{equation}
The Regge propagators $\Pi_v(s,t)$ depend on the $s$ variable 
as well as $t$ (see Eq. (B3) in Appendix B),
so we choose a fixed reference laboratory energy $E_{0}$ roughly 
corresponding to the minimum energy where Regge phenomenology 
is applicable. We take $E_0 = 4.0 \;$ GeV for this minimum energy, 
and since 
$s_0 = m_{p}(m_{p}+2E_0)$, we obtain $\sqrt{s_{0}} = 2.9 \;$ GeV. 
Finally, we fit the radiative decay constants of the $\rho$ and 
$\omega$ to the $\Gamma_{\rho \rightarrow \pi \gamma}$ and
$\Gamma_{\omega \rightarrow \pi \gamma}$ decay widths
yielding $g_{\rho \pi \gamma} = 0.75e/m_{\rho}$ and 
$g_{\omega \pi \gamma} = 1.82e/m_{\omega}$ with $e= 0.30282$.
 
We have estimated the $K^0 {\bar K^0}$ Born amplitude corresponding
to the double exchange of $K^*(892)$ and $\rho, \omega$ resonances.
Assuming that the coupling constant $g_{K^* K^0 \gamma} = -1.15e/m_{K^*}
= -0.389$ GeV$^{-1}$ from the known branching ratio of the $K^*(892)$
decay and $g_{K^* \rho K^0} = g_{K^* \omega K^0}/3 = -7.7$ GeV$^{-1}$
based on the application of vector meson dominance model, we find that
near the $K {\bar K}$ threshold the $K^+ K^-$ Born cross section
dominates over the $K^0 {\bar K^0}$ one. At the effective $K {\bar K}$
mass equal to 1 GeV, it is higher by more than a factor 3.
Thus in the numerical calculations we have neglected the neutral
kaon Born amplitude $V_{K^0 {\bar K^0}}$. For the $K^+ K^-$
photoproduction the Born amplitude $V_{K^0 {\bar K^0}}$ does not
contribute because of the cancellation between $t_{KK}^{I=1}$
and $t_{KK}^{I=0}$ as we discussed in the last Section (See Eq.(28)).
We also expect that the inclusion of $V_{K^0 {\bar K^0}}$ will not
change drastically our numerical results for the $\pi^+ \pi^-$ cross
sections. Certainly, our calculations can be refined in future
if the $K {\bar K}$ strong amplitudes are known better.

With all model parameters fixed according to the previous discussion,
we now present our numerical results. In all figures we plot only the 
$S$-wave projected cross sections and leave an investigation of 
interference with $P$-wave amplitudes for a future study. In Fig. 2(a)
we plot the invariant two pion mass ($M_{\pi \pi}$) dependence
of the Born cross section showing the sensitivity to choice of either
Bonn or Spanish $\rho NN$ parameters. 
We note that the $V_{I}$ amplitude 
(contribution from two exchanged charged pion graphs together with
the contact diagram)
dominates for small $M_{\pi \pi}$ 
and the double vector exchange amplitude $V_{II}$ dominates at higher 
two pion energy, $M_{\pi \pi} \stackrel{>}{\sim} 0.9 \;$ GeV.  
In Fig. 2(b) we show the effect of the purely on--shell 
final state interaction consecutively for single channel $\pi \pi$
intermediate states (dotted line above $K\bar{K}$ threshold and overlapping
solid line below) and the coupled $\pi \pi/K\bar{K}$ channels.
The Born cross section is presented
on the same graph for comparison. Fig. 2(c) shows the sensitivity to
the full FSI which includes both on--shell and off--shell 
principal value (PV) integrated amplitudes. Let us note that the 
off--shell $K\bar{K}$ amplitude makes a sizeable contribution
(eg. dip feature) below the physical $K\bar{K}$ threshold in
contrast with Fig. 2(b) where the on--shell $K\bar{K}$ channel 
has an effect only above threshold. The dramatic variation of the
cross section just below the $K\bar{K}$ threshold is due to the
$f_{0}(980)$ resonance in the coupled $\pi \pi \leftrightarrow K\bar{K}$
FSI amplitudes. 
An important point is that the $f_{0}(980)$ resonance shows up
as a peak in the $S$-wave two pion photoproduction cross 
section (enhanced by a factor three relative to the Born cross section), 
whereas in the elastic $S$-wave 
pion--pion scattering the $f_{0}(980)$ causes a dip in the cross section.
This happens because the unitary elastic amplitude, which behaves
like 
${t}_{\pi \pi} = \frac{1}{r_\pi} \sin \delta_{\pi \pi} 
\exp(i\delta_{\pi \pi})$ (where $\delta_{\pi \pi}$ is the $\pi \pi$
$I=0, S$-wave phase shift) goes to 
zero when 
$\delta_{\pi \pi} \rightarrow \pi$. 
This roughly corresponds to 
$M_{\pi \pi}$ approaching $M_{f_{0}} = 980 \;$ MeV 
as analyzed in Refs. \cite{klm,klr,kll}.
However, in two pion $I=0$ photoproduction the unitary amplitude has 
a purely real term associated with production without FSI, 
and the additional complex term related to the elastic scattering
in the final state,
hence
the full on--shell amplitude in Eq. (\ref{tpipi0}) below the \kk threshold 
behaves as 
$T_{\pi \pi}^{I=0} =
[1 + i \sin \delta_{\pi \pi} \exp(i\delta_{\pi \pi})] V_{\pi 
\pi}^{I=0} = \cos \delta_{\pi \pi} \exp(i \delta_{\pi \pi}) V_{\pi 
\pi}^{I=0}$, which produces a peak\
cross section when $\delta_{\pi \pi} = \pi$. 
This enhancement of the cross section in two pion photoproduction
at the $f_{0}(980)$ resonance makes direct measurements
of $f_{0}$ properties an interesting possibility at \tjnaf.                

In Figs. 3(a),(b) we show plots of the $t$-dependence of cross section
corresponding to calculations using $\rho, \omega$ normal and Regge  
propagator prescriptions, respectively.
In both cases we find a peak in the angular distribution for 
$|t|\approx 0.1-0.3$ GeV$^2$. 
A distinct feature of the Regge vector meson propagator is 
a node at $t_{n} \approx -0.5 \;$ GeV$^2$ which arises due to its analytic
dependence on the linear Regge trajectory.
The $\rho$ and $\omega$ trajectories 
pass through zero at $t_n$ causing a zero in
the Regge propagator (see Eq.(\ref{props})).
Besides the obvious difference in the shape of the $t$-dependence,
the cross section calculated using normal and Regge propagators 
significantly differ in their magnitudes.      

In Figs. 4(a),(b) we demonstrate sensitivity to the principal value (PV)  
form factor prescription (i.e. choice of $F_{cut}(M',M)$ in Eq.(\ref{vhat})) 
and cut-off parameter $\Lambda_{cut}$. 
We use cut-off functions decreasing asymptotically as $(M')^{-4}$ 
because it is the minimum power required to give convergence for all 
amplitudes considered.
In Fig. 4(a), we show the
$M_{\pi \pi}$ distribution calculating the PV 
integrals with a cut-off form factor that gives 
suppression for all off shell $M'$ values,
\begin{equation}
F_{cut}(M',M) = \left(\frac{\Lambda_{cut}^2}
{\Lambda_{cut}^2 + (M'-M)^2} \right)^2 .
\label{dipole}
\end{equation}
In Fig. 4(b), we used a cut-off form factor that gives 
enhancement for off shell $M' < M$ and suppression for $M'>M$,
\begin{equation}
F_{cut}(M',M) = \left(\frac{\Lambda_{cut}^2 + M^2}
{\Lambda_{cut}^2 + M'^2} \right)^2 .
\label{power}
\end{equation}
Of course, the use of a global form factor is not unique in our work.
S. Nozawa et al.~\cite{nozawa} used the global form factor
as a function of the momentum. Since we can easily make a relation of the
effective mass to the c.m. momentum, our method is very similar to
S. Nozawa's et al. The form given by Eq.(32) is also related to the
off-shell behaviour of separable amplitudes. 
As we can see from Figs. 4(a),(b), the differential cross sections at small 
$M_{\pi\pi}$ ($<$ 0.4 GeV)
are not sensitive to the form of $F_{cut}$ function. As $M_{\pi\pi}$
gets larger, the dependence on the $F_{cut}$ function and its parameter $\Lambda_{cut}$
become significant.
However, the general structure of peaks and dips remains same and especially 
the peak around 1 GeV corresponding to $f_0(980)$ is not changed.
Let us note that the contribution of the principal value part 
(or the off--shell part) to the photoproduction amplitude is sizable and 
cannot be neglected in comparison with the on--shell part.
In all figures, including Figs. 2 and 3, we 
use the cut-off function given by Eq. (\ref{dipole}) with $\Lambda_{cut} = 
1.0 \;$ GeV and Bonn parameters unless otherwise specified.

In Figs. 5, 6 and 7 we present our results for the $K^{+} K^{-}$ final state
channel. In Fig. 5(a) we plot the Born cross
section $M_{KK}$ dependence, showing 
the dependence on the $\rho N N$ parameters
$G_v^{\rho}$ and $G_T^{\rho}$. 
Figs. 5(b),(c) compare the $M_{KK}$ on--shell FSI
distribution relative to the
Born cross section. 
Note that the on shell FSI effect suppresses the Born cross section
for both the single and coupled channel results. In Fig. 5(d) we 
show the effect of the full final state interactions.
Now the coupled channel FSI gives a substantial enhancement
relative to the Born cross section just above the $K\bar{K}$
threshold that is absent in the result for the single channel FSI.       
Fig. 5(e) shows an expanded $M_{KK}$ range of Fig. 5(d). 
The sharp decrease in the cross section near $M_{KK} \approx 1.4 \;$ GeV
arises due to the $f_{0}(1400)$ resonance in the FSI~\cite{klm,klr,kll}.

In Fig. 6 we present the t-dependence of the M-integrated 
$K^{+}K^{-}$ cross section 
\be
d\sigma/dt = \int_{2m_K}^{M_{cut}} \frac{d\sigma}{dtdM} dM
\label{integral}
\ee
showing results obtained using normal and Regge propagators. 
Experimental points represent the differential cross section for
the elastic $K^+ K^-$ photoproduction in the $\phi$ meson effective
mass range and for the photon energy range $3.8 < E_{\gamma} < 4.8$
GeV~\cite{barberetal}. The data show the forward diffractive peak 
corresponding to the dominant $P$-wave $K^+ K^-$ production.
At higher $-t$, where we observe a change of the cross section slope,
one may expect some contribution from the $S$-wave $K^+ K^-$ production.
Indeed, the calculated $S$-wave cross sections given by solid
or dashed lines are more comparable with experimental data
for higher $-t$ region.
Again as in the \pp photoproduction, a characteristic minimum at $t \approx -0.5$ GeV$^2$ for the 
calculations performed with Regge $\rho$ and $\omega$ propagators
is a result of the zeroes of the Regge trajectories. 
Final state interactions increase the Born $t-$distributions in all cases.
This effect is amplified at higher $t$.

Figs. 7(a),(b) show the $M_{KK}$-dependence of the t-integrated
$K^{+}K^{-}$  cross section
\be
d\sigma/dM = \int_{t_{cut}}^{t_{min}} \frac{d\sigma}{dtdM} dt.
\label{integral2}
\ee
The minimum $|t|$-value, $|t_{min}|$, is an increasing function of 
$M_{KK}$.   In Fig. 7(a) we take the t-integration range 
-1.5 GeV$^2$ $\leq t \leq t_{min}$
corresponding to the data of
Ref.~\cite{barberetal}, whereas in Fig. 7(b) we integrate
over the range 
-0.2 GeV$^2$ $\leq t \leq$ $t_{min}$
corresponding to the data of Ref.~\cite{Friesetal}.
In Table 2 we also present the $S$-wave \kk total photoproduction cross section 
integrated over the effective \kk mass from threshold up to 1.04 GeV and
over two ranges of $t$.
Important difference between cross section values for normal and Regge propagators
is related to different $t-$dependence (see Fig. (6)).
A few experimental data of the total cross section
for the pion and kaon pair productions were published
in various energy ranges\cite{barberetal,Friesetal,behrend,aston}.
Among them, the most relevant data to our calculations in this work
may be found in Ref. \cite{barberetal} where the photon energy was 
between 2.8 and 4.8 GeV and $|t|$ range was wide to cover even up to 
1.5 GeV$^2$. In the DESY experiment \cite{Friesetal} the energy range
was higher, between 4.6 GeV and 6.7 GeV.
The analysis of Ref. \cite{barberetal} showed that the $S$-wave cross-section,
assuming $f_0(980)$ resonance photoproduction, was $96.2 \pm 20.0$ nb and the 
cross-section, under the hypothesis that the \kk $S$-wave production
is nonresonant, was $10.2 \pm 4.1$ nb. 
Even though our prediction does depend on
the detailed theoretical input, our values of total cross section
over $M_{KK}$ between 0.99 and 1.04 GeV and $t$ above $-1.5$ GeV$^2$
are of the same order as the 
available data \cite{barberetal,Friesetal,behrend}.
Using the Regge propagators for $\rho$ and $\omega$ mesons, 
our values were reduced by an
order of magnitude compare to the values with the normal propagators.
Theoretically, it seems to us more reasonable
to use the Regge propagator rather than the normal propagator for the
$\rho$ and $\omega$ mesons in the high photon energy region. 

The integrated cross sections quoted  in
\cite{barberetal,Friesetal} 
differ by more than order of magnitude. 
A clear difference is also seen if we compare Figs. 7(a) and 7(b)
calculated at the same energy but in the different $t-$ranges.
The difference between the experimental cross sections could also be related 
to the energy dependence.
In Figs. 8(a),(b) the photon energy dependence of $\frac{d\sigma}{dtdM_{KK}}$
is given for $M_{KK} = 1$ GeV and two values of $t$. 
In Figs. 8(a), the normal propagators
of $\omega, \rho$ are used and in Fig. 8(b), the Regge propagators
are used. As we can see from Figs. 8(a) and 8(b), the photon energy
dependence is drastically different depending on the choice
of propagators. The $\rho, \omega$ Regge exchanges lead to the cross sections 
decreasing sharply with increasing photon energy while the normal propagators 
give the opposite behaviour of the $E_{\gamma}$ dependence.

\section{Conclusions}

              
Our calculations indicate that the final state interactions 
are crucial to determine the structure of differential cross sections.
The Born effective mass distributions are structureless
while the final state interactions produce dips and peaks near the resonances 
(see Figs. 2(c), 4(a), 4(b) and 5(e)).
Our calculations include $\pi \pi$ and $K {\bar K}$ coupled
channels as well as the on-shell and off-shell
contributions.
Especially, the contributions from the scalar resonances, $f_0(980)$
and $f_0(1400)$, are evident in the effective mass distributions.

We have calculated the differential cross sections as functions of the effective 
masses and momentum transfers. The total cross sections are shown in Table 2.
Our predictions of the total cross sections indicate that the $\pi^+\pi^-$ 
and $K^+K^-$
photoproduction processes are experimentally measurable in the photon energy
range of a few GeV.
The Regge predictions for the total cross section depend on the normalization of 
the Regge vertex functions.
If we increase the value of $s_0$ used in the normalization of the Regge vertex
functions, then our predictions of the total cross sections become even larger.
For example, if we increase our $\sqrt{s_0}$ by 10$\%$, then the cross section
increases by 25$\%$. 

Since the calculation in this work is limited only to the $S$-wave
partial amplitude in the final states, we do not discuss the $P$-wave
interference effect and the contribution of the $\phi$-meson
to the \kpkm photoproduction. The $t$-distribution of $K^+ K^-$
production in Fig. 6 presents the typical $S$-wave contribution which 
has the minimum in the forward direction followed by the maximum
at a slightly larger $|t|$ value. On the other hand, the diffractive
$\phi$-production has the peak at the forward direction\cite{barberetal}.
To see the interference effect between $S$ and $P$ wave contributions,
one should not limit the range of $|t|$ below 0.2 GeV$^2$
but go up to higher values such as 1.5 GeV$^2$.

We have tried to establish a baseline of comparison to help motivate
interest in new photo- and electroproduction experiments. By careful
experimental study in small bins of $\pi \pi$ and $K {\bar K}$ masses,
one can obtain valuable information on the positions and widths of scalar
resonances, especially the $f_0(980)$ meson. Based on our calculations,
precision two-pion/kaon data will be extremely valuable for constraining
the effective off-shell behavior of the elementary production amplitudes.
Additionally, the accurate measurements of the neutral pion pairs
$\pi^0 \pi^0$ and $K^0 {\bar K}^0$ photoproduction will be crucial to
separate the different isospin contributions of $I=0,1$ and 2 states.

\vspace{0.7in}
\centerline{\bf Acknowledgments}

\noindent

We are grateful to H. Funsten, G. Gilfoyle, N. Isgur, B. Kerbikov 
and B. Niczyporuk for
valuable discussions.
This work has been partially supported by the Polish State 
Committee for Scientific Research (grant No 2 P03B231 08), by the Maria
Sk\l{}odowska--Curie Fund II (No PAA/NSF--94--158),  
National Science Foundation
under the international joint research (INT-9514904) and also in part by the
U.S. Department of Energy (DE-FG02-96ER40947). 
The North Carolina Supercomputer Center
and the National Energy Research Scientific Computing Center
are also acknowledged for the grant of computing time
allocation.

\newpage

\renewcommand{\theequation}{A\arabic{equation}}
\renewcommand{\thesection}{}
 
\section*{Appendix A}
\setcounter{equation}{0}

\begin{center}
{\bf Effective interaction Lagrangian}
\end{center}

\vspace{0.3cm}

The effective interaction Lagrangian density used in this work is 
summarized as follows:
\begin{eqnarray}
{\cal L} &=& {\cal L}_{KK\gamma} + {\cal L}_{\pi\pi\gamma} + 
{\cal L}_{\rho\pi\pi} + {\cal L}_{\rho\pi\pi\gamma} + {\cal L}_{\rho K K \gamma} + 
{\cal L}_{\rho KK} + \nonumber\\
&&{\cal L}_{\omega KK} + 
 {\cal L}_{\rho\pi\gamma} + 
{\cal L}_{\omega\pi\gamma} + {\cal L}_{\rho \pi\omega} +
{\cal L}_{\rho NN} + {\cal L}_{\omega NN},
\end{eqnarray}
where
\begin{eqnarray}
{\cal L}_{KK\gamma} &=& ie(\partial^\mu K^\dagger\frac{1 + \tau_3}{2}K 
- K^\dagger\frac{1+\tau_3}{2}\partial^{\mu}K)A_{\mu},
\\
{\cal L}_{\pi\pi\gamma} &=& - e\epsilon_{3ij}\pi^i\partial^\mu\pi^jA_\mu, 
\\ 
{\cal L}_{\rho\pi\pi} &=& 
-g_{\rho \pi \pi}\epsilon_{ijk}\pi^i\partial^\mu\pi^j\rho_\mu^k,
\\ 
{\cal L}_{\rho\pi\pi\gamma} &=& 
-eg_{\rho \pi \pi}\epsilon_{ijk}\epsilon_{3lk}\rho_\mu^i\pi^j\pi^lA^\mu,
\\
{\cal L}_{\rho K K \gamma} &=& 
e g_{\rho KK}K^\dagger(\vec\tau\cdot\vec\rho_\mu + \rho_{\mu}^3)K A^{\mu},
\\
{\cal L}_{\rho KK} &=& 
ig_{\rho KK} (\partial^\mu K^\dagger \vec\tau\cdot\vec\rho_\mu K -
K^\dagger {\vec \tau}\cdot{\vec \rho}_\mu \partial^\mu K), 
\\
{\cal L}_{\omega KK} &=& 
ig_{\omega K K }(\partial^\mu K^\dagger K - K^\dagger \partial^{\mu}K)
\omega_{\mu},
\\
{\cal L}_{\rho\pi\gamma} &=& 
g_{\rho \pi \gamma} \epsilon^{\mu \nu \lambda \sigma} \partial_{\mu}
A_{\nu}{\vec \pi}\cdot\partial_{\lambda}{\vec \rho}_{\sigma},
\\
{\cal L}_{\omega\pi\gamma} &=&
g_{\omega \pi \gamma} \epsilon^{\mu \nu \lambda \sigma}
\partial_{\mu} A_{\nu} \partial_{\lambda}
\omega_\sigma \pi^i \delta_{i3},
\\
{\cal L}_{\rho \pi\omega} &=&
g_{\rho \pi \omega} \epsilon^{\mu \nu \lambda \sigma} \partial_{\mu}
\omega_{\nu}{\vec \pi} \cdot  \partial_{\lambda}{\vec \rho}_\sigma,
\\
{\cal L}_{\rho NN} &=& 
-{\bar \psi}(G_{\rho}^{V}\gamma^{\mu}
-\frac{G_{\rho}^T}{2 m_N}\sigma^{\mu \nu}\partial_{\nu})
\vec\tau\cdot\vec\rho_\mu \psi,
\\
{\cal L}_{\omega NN} &=& 
- {\bar \psi}(G_{\omega}^{V}\gamma^{\mu}
- \frac{G_{\omega}^T}{2 m_N}\sigma^{\mu \nu}\partial_{\nu})
\omega_{\mu}\psi.
\end{eqnarray}

The pion fields ${\vec \pi}=(\pi^1,\pi^2,\pi^3)$ are
given by 
\begin{eqnarray}
\pi^1 &=& \frac{1}{\sqrt{2}}(\pi^- - \pi^+)\\
\nonumber
\pi^2 &=& \frac{i}{\sqrt{2}}(\pi^+ + \pi^-)\\
\nonumber
\pi^3 &=& \pi^0,
\end{eqnarray}
 with $\pi^{\pm}$ being the creation operators of $\pi^{\pm}$ states 
{\it i.e.} 
 $\pi^{\pm}|0\rangle = |\pi^\pm\rangle$. 
 The $\rho$ fields ${\vec \rho}=(\rho^1,
\rho^2,\rho^3)$ are defined similarly.
Also, the kaon doublets are given by
$K = (K^+, K^0)$, $K^+ = (K^-, \bar{K}^0)$. 
The photon field and the $\omega$ meson field
are denoted as $A^\mu$ and $\omega^\mu$.
The nucleon doublet field is given by $\psi$ and $\tau$ is the 
isospin-$\frac{1}{2}$ operator.

\newpage

\renewcommand{\theequation}{B\arabic{equation}}
\renewcommand{\thesection}{}
 
\section*{Appendix B}
\setcounter{equation}{0}


\begin{center}
{\bf Detailed factors in the current $J_{r,m\bar m}$}
\end{center}

\vspace{0.3cm}

The variables used in $J_{r,m\bar m}$ (Eq. 4), {\it i.e.} 
$\alpha_{r,m\bar m}, \beta_{1r,m \bar m}, \beta_{2r,m\bar m}, d_{r,m\bar m}$ 
and $e_{r,m\bar m}$ are given by
\begin{eqnarray}
&& \alpha_{I,m\bar m} = 2,  \\ \nonumber
&&\alpha_{II,m\bar m} = \frac{(q\cdot k_1)(k_1 \cdot k_2) + 
(q \cdot k_1)(q \cdot k_2)
- m_m^2(q \cdot k_2)}{m_m^2 - m_a^2 - 2{q \cdot k_1}}
+ (k_1 \leftrightarrow k_2), \\ \nonumber
&&{\beta_1}_{I,m\bar m} = -\frac{q-k_1+k_2}{q \cdot k_1}, \\ \nonumber
&&{\beta_2}_{I,m\bar m} = (\beta_1)_I (k_1 \leftrightarrow k_2), \\ \nonumber
&&{\beta_1}_{II,m\bar m} = \frac{(q\cdot k_2) k_1 - (k_1 \cdot k_2) q}
{m_{m}^2 - m_{a}^2 - 2 {q \cdot k_1}}
+ \frac{(m_{m}^2 - {q\cdot k_2}) q - (q\cdot k_2) k_2}
{m_{m}^2 - m_{a}^2 - 2 {q\cdot k_2}}, \\ \nonumber
&&{\beta_2}_{II,m\bar m} = (\beta_1)_{II,m\bar m} (k_1 \leftrightarrow k_2), \\ \nonumber
&&d_{I,m\bar m} = e g_{abm} (G_V^{b} + G_T^{b}) \; \Pi_{b}(s,t)\;F_{bNN}(t),
\\ \nonumber
&&e_{I,m\bar m} =
 -e g_{abm} (\frac{G_T^{b}}{2m_p}) \; \Pi_{b}(s,t)\;F_{bNN}(t), \\
\nonumber
&&d_{II,m\bar m} = g_{abm} g_{am\gamma}
(G_V^{b}+G_T^{b}) \; \Pi_{b}(s,t)\;F_{bNN}(t), \\ \nonumber
&&e_{II,m\bar m} =  -g_{abm} g_{am\gamma}
(\frac{G_T^{b}}{2m_p}) \; \Pi_{b}(s,t)\;F_{bNN}(t).
\label{variables}
\end{eqnarray}

Here $m_p,m_{m},m_a$ and $m_b$ denote the masses of the proton and
the mesons $m,a$ and $b$ in Fig. 1, respectively.
Also, the subscripts of coupling constant $g$ denote the mesons
involved in the interaction vertex.
For example, for the case of $m\bar m = \pi^+ \pi^-$,
the couplings of $\rho^0 \pi^+ \pi^-$, $\omega \rho^0 \pi^0$
and $\rho^0 \pi^0 \gamma$ are denoted by $g_{\rho \pi \pi}$,
$g_{\omega \rho \pi}$ and $g_{\rho \pi \gamma}$, respectively, and
the vector and tensor couplings of $\rho(\omega)$ are given by
$G_V^{\rho}(G_V^{\omega})$ and $G_T^{\rho}(G_T^{\omega})$, respectively.
We denote the $\rho$ and $\omega$ propagators by $\Pi_{b}(s,t)$
($b=\rho ,\omega$) and present results for both normal (free) and
Regge propagators given by the expressions:
\begin{eqnarray}
\Pi_{b}(s,t) &=& \frac{1}{t - m_{v}^2} \;\;\; (normal), \\
\Pi_{b}(s,t) &=& \frac{-1}{2s^{\alpha_0}}(1 - e^{i\pi \alpha_{b}(t)}) \;
\Gamma(1-\alpha_{b}(t)) \;
(\alpha' s)^{\alpha_{b}(t)} \;\;\; (Regge).
\label{props}
\end{eqnarray}
The vector meson propagators are expressed in terms of the Mandelstam
variables $s = (q + p)^2$, and $t = (p'-p)^2$, where $q,\; p,\; p'$ are
the photon, initial proton, and final proton 4-momentum, respectively.
The $\rho, \omega$ Regge propagators also depend on the vector meson
trajectories which are
parameterized by $\alpha_{b}(t) = \alpha_{0} + \alpha'(t - m_{b}^2)$,
with $\alpha_{0} = 1.0$ and $\alpha' = 0.9 \;$ GeV$^{-2}$ \cite{Irving}.
The function $F_{bNN}(t)$ is the meson $b$-nucleon form factor.

\newpage

\renewcommand{\theequation}{C\arabic{equation}}
\renewcommand{\thesection}{}
 
\section*{Appendix C}
\setcounter{equation}{0}

\begin{center}
{\bf $S$-wave amplitude}
\end{center}
In this Appendix, the $S$-wave ($l = 0, l_z = 0$)
results of the integration over the solid angle
 in Eq.(\ref{stensor}) are summarized. For $r = I$, we find that 
\begin{equation}
P_I^{\mu \nu} = 2 (g^{\mu \nu} - h^{\mu \nu}),
\label{pimunu}
\end{equation}
where
\begin{eqnarray}
h^{00} &=& xQ_0(x),\nonumber \\
h^{0i} &=& x\{Q_0(x)-2\frac{|{\vec\kappa}|}
{|{\vec q}|}Q_1(x)\}{\hat q_i},\nonumber \\
h^{i0} &=& Q_1(x){\hat q_i},\nonumber \\
h^{ij} &=& -\frac{2|{\vec \kappa}|}{3|{\vec q}|}
\{Q_0(x) - Q_2(x)\}\delta_{ij} \nonumber \\
&+&\{Q_1(x)-2\frac{|{\vec \kappa}|}{|{\vec q}|}Q_2(x)\}{\hat q_i}{\hat q_j}
\;\;\;\;\; (i,j=1,2,3).
\end{eqnarray}
Here $x=\sqrt{1+{m_m^2}/{|\vec{\kappa}|^2}}$,
$|{\vec q}| = (M_{m m}^2 - t)/(2 M_{m m})$
is the photon momentum in the $m \bar m$ c.m. system
and ${\hat q}_i$ are the photon momentum coordinates
normalized to 1 ($\sum_i {\hat q}_i^2 = 1$).
The Legendre functions of the second kind $Q_0,Q_1$,$Q_2$ are
given by
\begin{eqnarray}
Q_0 (x) &=& \frac{1}{2} ln\frac{x+1}{x-1},\nonumber \\
Q_1 (x) &=& \frac{x}{2} ln\frac{x+1}{x-1} - 1,\nonumber \\
Q_2 (x) &=& \frac{1}{4} (3x^2-1)ln\frac{x+1}{x-1} - \frac{3}{2} x.
\end{eqnarray}

Similarly we obtain $P_{II}^{\mu \nu}$ as follows;
\begin{eqnarray}
P_{II}^{00} &=& \frac{1}{3}|\vec{\kappa}||\vec{q}|
[Q_0(y) -3xQ_1(y) + 2 Q_2(y)],\nonumber \\
P_{II}^{i0} &=& \frac{1}{3}|\vec{\kappa}||\vec{q}|{\hat q_i}
[Q_0(y) -3xQ_1(y) + 2 Q_2(y)],\nonumber \\
P_{II}^{0i} &=& \frac{1}{3}|\vec{\kappa}|^2{\hat q_i}x
[4\{Q_0(y) -Q_2(y)\}+3\frac{|{\vec q}|}{|\vec {\kappa}|}
\{xQ_0(y) - Q_1(y)\}],\nonumber \\
P_{II}^{ij} &=& \frac{1}{3}|\vec{\kappa}|^2 [2x\{2Q_0(y)
-3xQ_1(y)+Q_2(y)\}+\frac{|{\vec q}|}{|\vec{\kappa}|}\{
(3x^2-1)Q_0(y)-2Q_2(y)\}]\delta_{ij} \nonumber \\
&+&{\hat q_i}{\hat q_j} \frac{1}{3}|\vec{\kappa}|^2 [6x\{xQ_1(y)
-Q_2(y)\} + \frac{|{\vec q}|}{|\vec{\kappa}|}
\{Q_0(y) -3xQ_1(y) + 2 Q_2(y)\}],
\end{eqnarray}
where
\begin{equation}
y = x + \frac{m_a^2 - m_m^2}{2|{\vec q}||{\vec{\kappa}}|}.
\end{equation}
For the $K^+ K^-$ photoproduction, we need only $P_{I}^{\mu \nu}$
given by Eq.~(\ref{pimunu}).



\newpage

\begin{table}[htb]
\centering
\caption{Summary of calculated diagrams} 
\vspace{0.4cm}
 \begin{tabular}{ccc}
\hline
 $m{\bar m}$ & $r = I$ (three diagrams) & $r = II$ (two diagrams) \\ 
\hline
 $\pi^+\pi^-$ & $(a,b) = (\pi^{\pm},\rho^0)$ & $(a,b) = (\rho^{\pm},\omega)$ \\ 
 $\pi^0\pi^0$ & --------------  & $(\rho^0,\omega)$,$(\omega,\rho^0)$ \\
 $K^+K^-$ & $(K^{\pm},\rho^0)$,$(K^{\pm},\omega)$ & --------------\\
\hline
\end{tabular}
\end{table}

\vspace{1cm}

\begin{table}[htb]
\centering
\caption{Born and full (with all FSI) $S$-wave \kk total photoproduction 
cross sections for normal and Regge propagators} 
\vspace{0.4cm}
 \begin{tabular}{ccc}
\hline
& \multicolumn{2}{c}{cross section (nbarn) for:} \\
\cline{2-3}
amplitudes, propagator & $|t| \leq 1.5$ GeV$^2$ & $|t| \leq 0.2$ GeV$^2$ \\
\hline
Born, normal & 121 & 20 \\
full, normal & 464 & 46 \\
Born, Regge &  12 & 8 \\
full, Regge & 35 & 18 \\
\hline
\end{tabular}
\end{table}

\newpage

\begin{figure}[htb]
\caption{Generic structure of diagrams contributing to \mm photoproduction;
$a$ and $b$ stand for $\rho$, $\omega$ or $\pi$ mesons and \mm denotes pairs
of \pp, \pnpn or \kk.}

\vspace{0.5cm}

\xslide{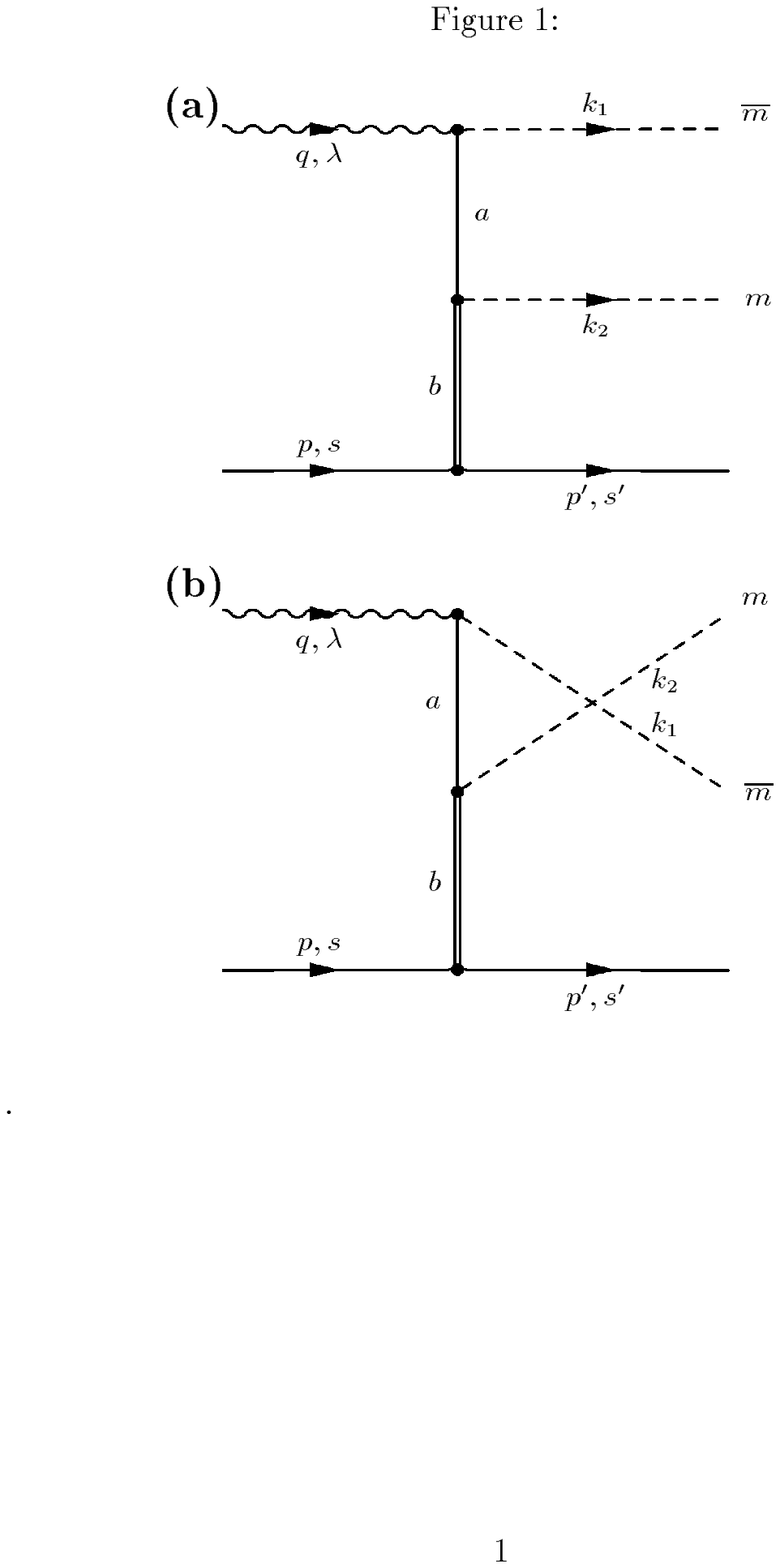}{14.5cm}{180}{335}{405}{672}{8cm} 

\label{fig1}
\end{figure}

\renewcommand{\thefigure}{2(a)}

\begin{figure}[htb]
\caption{
Invariant mass distribution for $S$-wave $\pi^{+} \pi^{-}$ 
photoproduction at 
$E_{\gamma}^{lab} = 4.0 \;$ GeV and $t = -0.2 \;$ GeV$^2$. All curves
are calculated without final state interactions.  
The solid and dotted lines 
are calculated with the full Born amplitude using Bonn and Spanish
parameters respectively. The dashed line shows the contribution from
the $V_{I}$ amplitude (i.e. without double $\rho + \omega$ exchanges) using
Bonn parameters. The dotted--dashed line corresponds to $\rho + \omega$ 
exchanges ($V_{II}$ amplitude).
}

\vspace{0.5cm}

\xslide{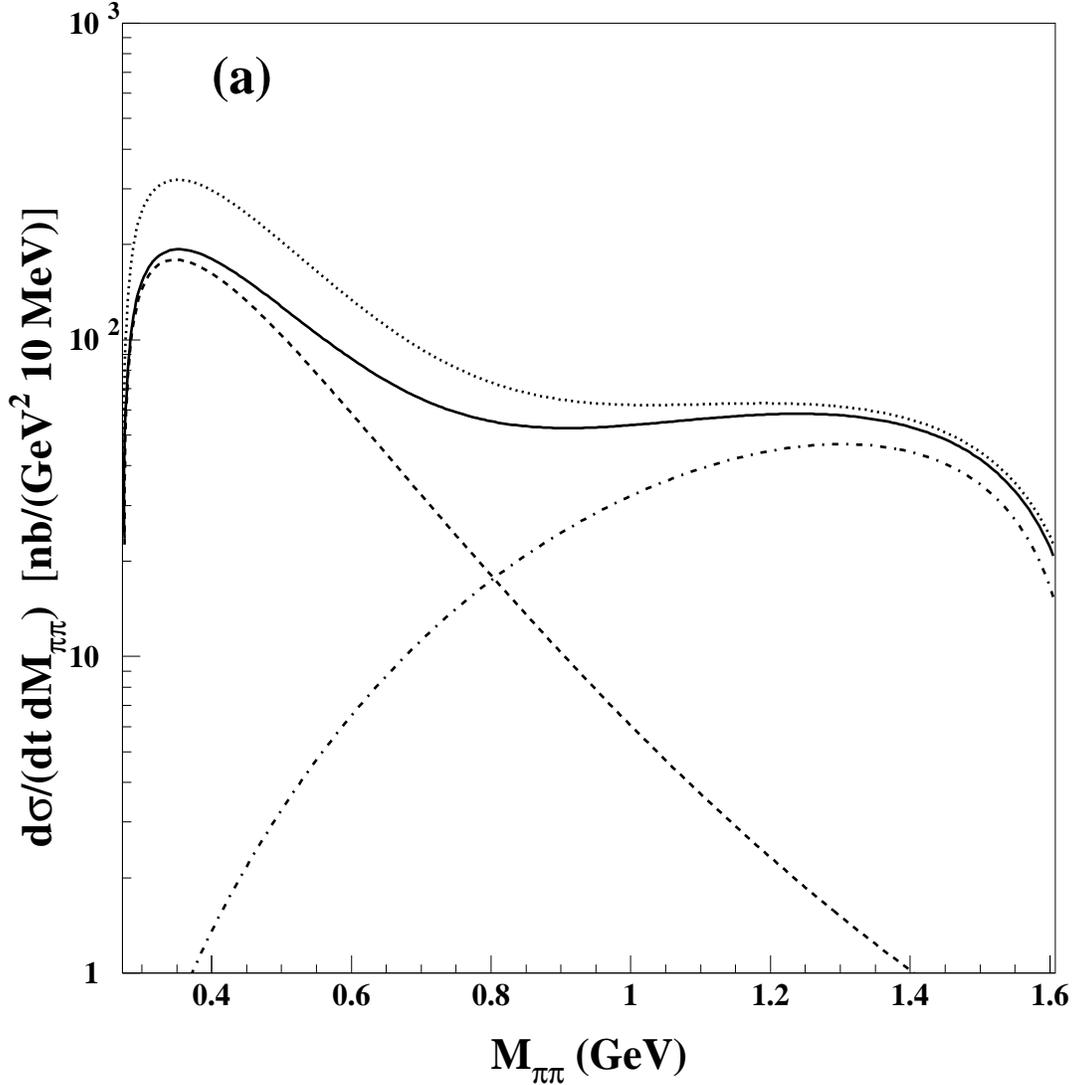}{14.5cm}{10}{145}{530}{655}{14.5cm} 

\label{fig2a}
\end{figure}

\renewcommand{\thefigure}{2(b)}

\begin{figure}[htb]
\caption{$S$-wave $\pi^{+} \pi^{-}$ invariant mass distribution  at 
$E_{\gamma}^{lab} = 4.0 \;$ GeV and $t = -0.2 \;$ GeV$^2$ showing 
sensitivity to purely on--shell  FSI terms
relative to the Born cross section (dashed line). The solid line 
shows the on--shell FSI result with both $\pi \pi$ and $K\bar{K}$ 
intermediate channels whereas the dotted line represents the result with
no $K\bar{K}$ coupling.}

\vspace{0.5cm}

\xslide{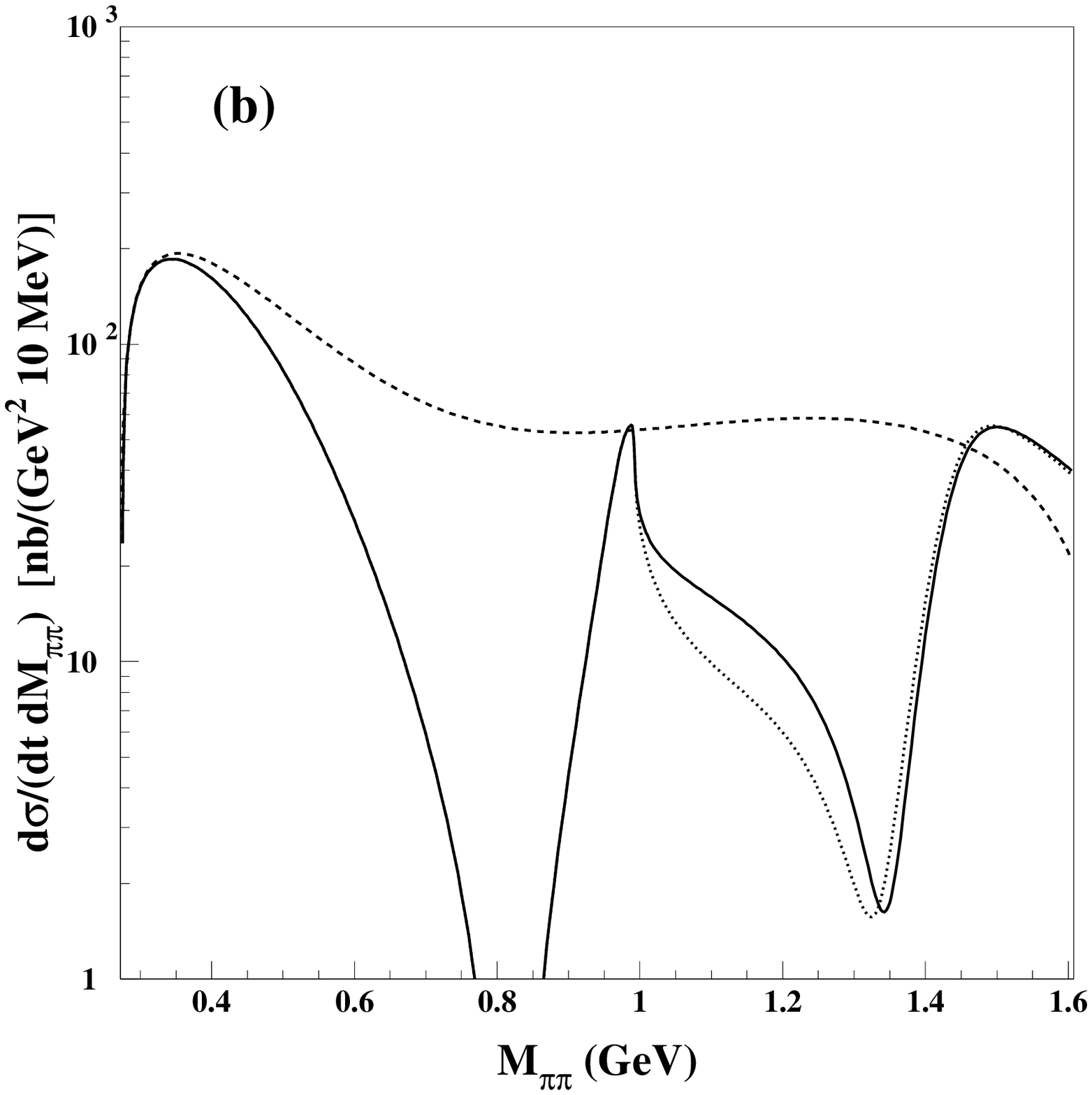}{14.5cm}{10}{145}{530}{655}{14.5cm} 

\label{fig2b}
\end{figure}

\renewcommand{\thefigure}{2(c)}
\begin{figure}
\caption{$S$-wave $\pi^{+} \pi^{-}$ invariant mass distribution at 
$E_{\gamma}^{lab} = 4.0 \;$ GeV and $t = -0.2 \;$ GeV$^2$. 
The solid line 
shows the full FSI result (on--shell and off--shell) with both $\pi \pi$ and $K\bar{K}$ 
intermediate channels, the dotted line represents the result with
no $K\bar{K}$ coupling and the dashed line corresponds to the Born cross section.}

\vspace{0.5cm}

\xslide{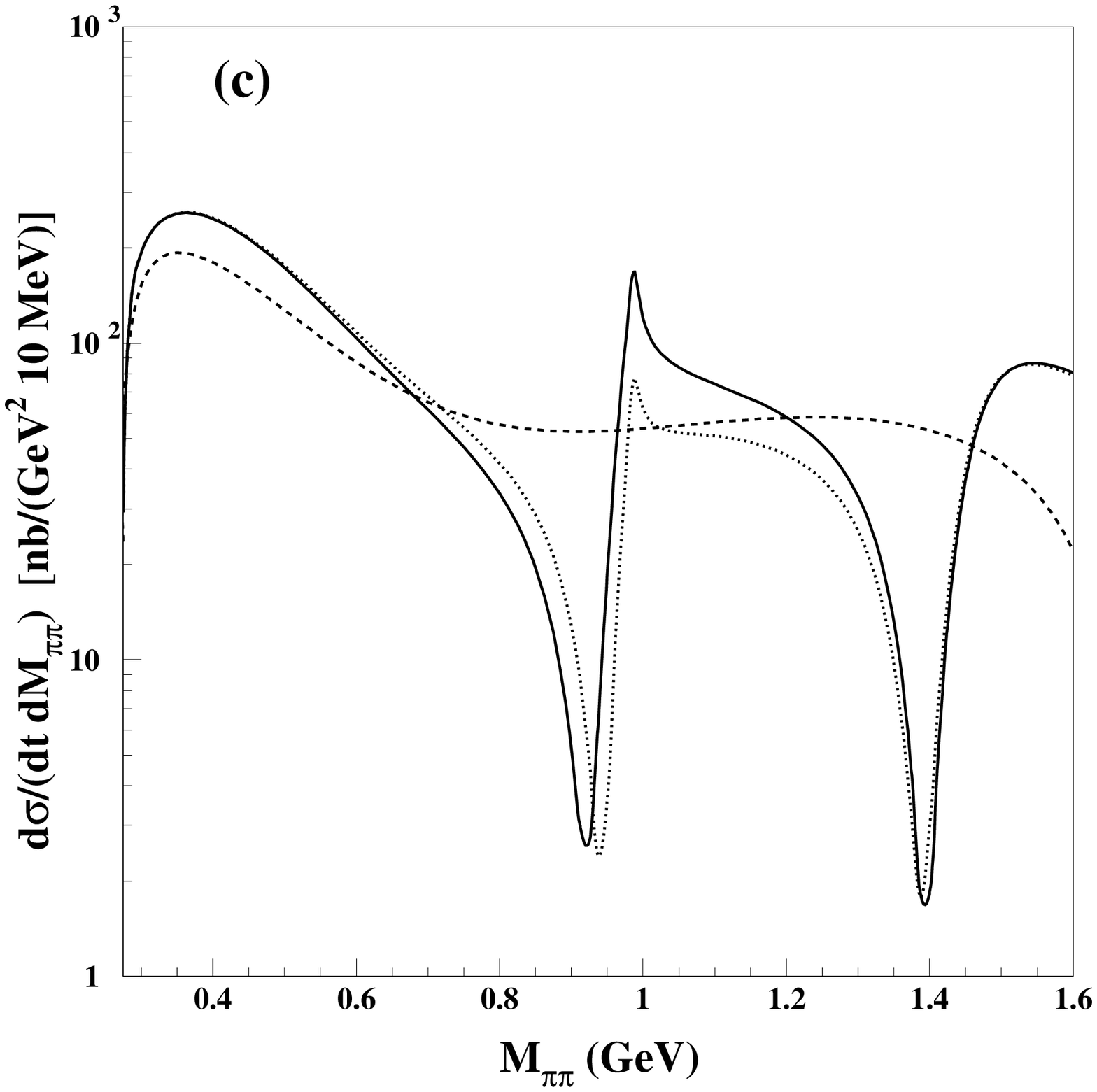}{14.5cm}{10}{145}{530}{655}{14.5cm} 

\label{fig2c}
\end{figure}

%
%
\renewcommand{\thefigure}{3(a)}
\begin{figure}
\caption{Invariant momentum transfer squared distribution 
for $S$-wave 
$\pi^{+} \pi^{-}$ photoproduction at $E_{\gamma}^{lab} = 4.0 \;$ GeV using
normal propagators.
The solid, dashed, and dotted lines were calculated with
$M_{\pi \pi}$ = 0.4, 0.7, and 1.0 GeV, respectively.}

\vspace{0.5cm}

\xslide{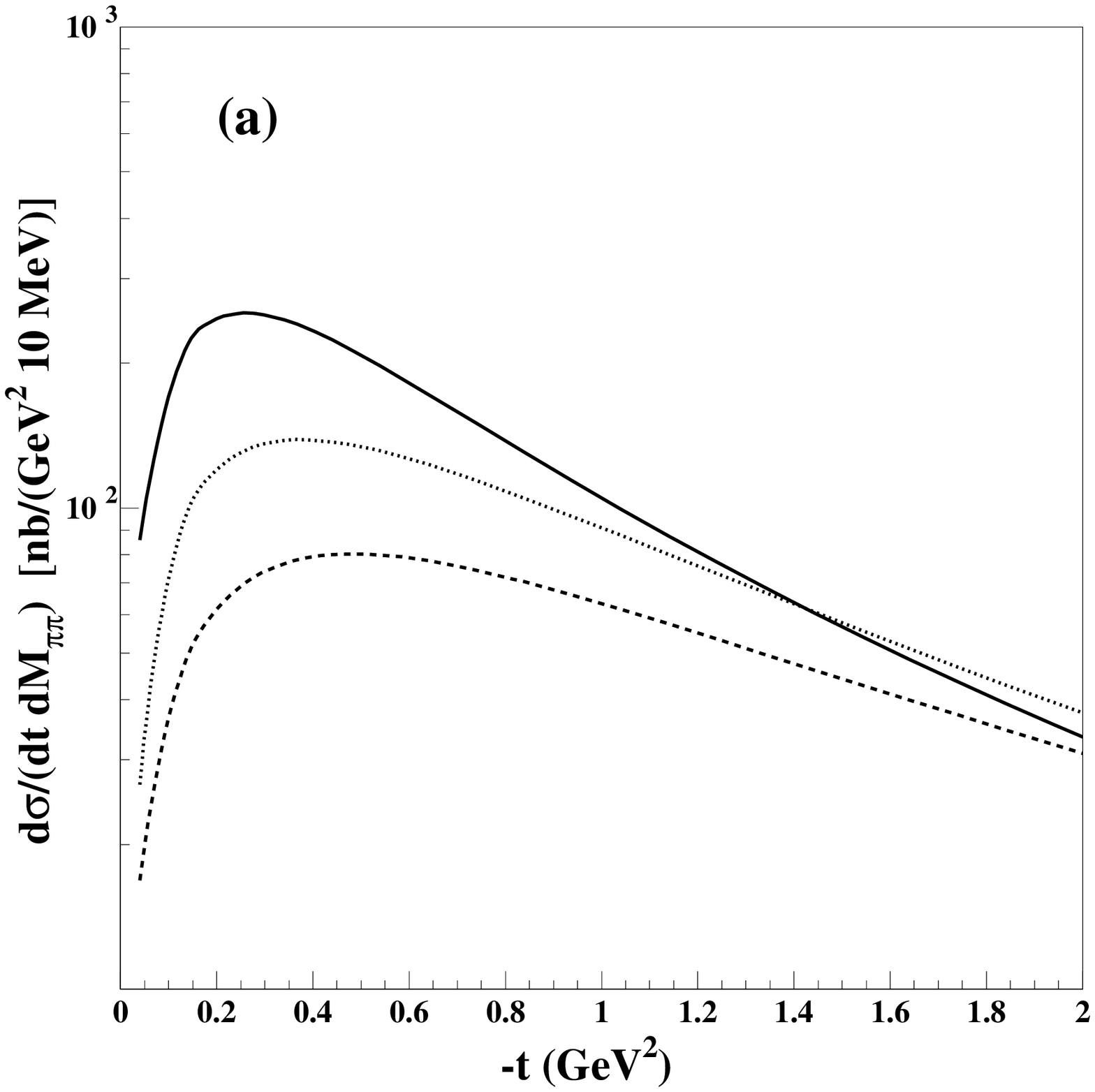}{14.5cm}{10}{145}{530}{655}{14.5cm} 

\label{fig3a}
\end{figure}

\renewcommand{\thefigure}{3(b)}
\begin{figure}
\caption{Invariant momentum transfer squared distribution for $S$-wave 
$\pi^{+} \pi^{-}$ photoproduction at $E_{\gamma}^{lab} = 4.0 \;$ GeV using
Regge propagators.
The solid, dashed, and dotted lines were calculated with
$M_{\pi \pi}$ = 0.4, 0.7, and 1.0 GeV, respectively.}

\vspace{0.5cm}

\xslide{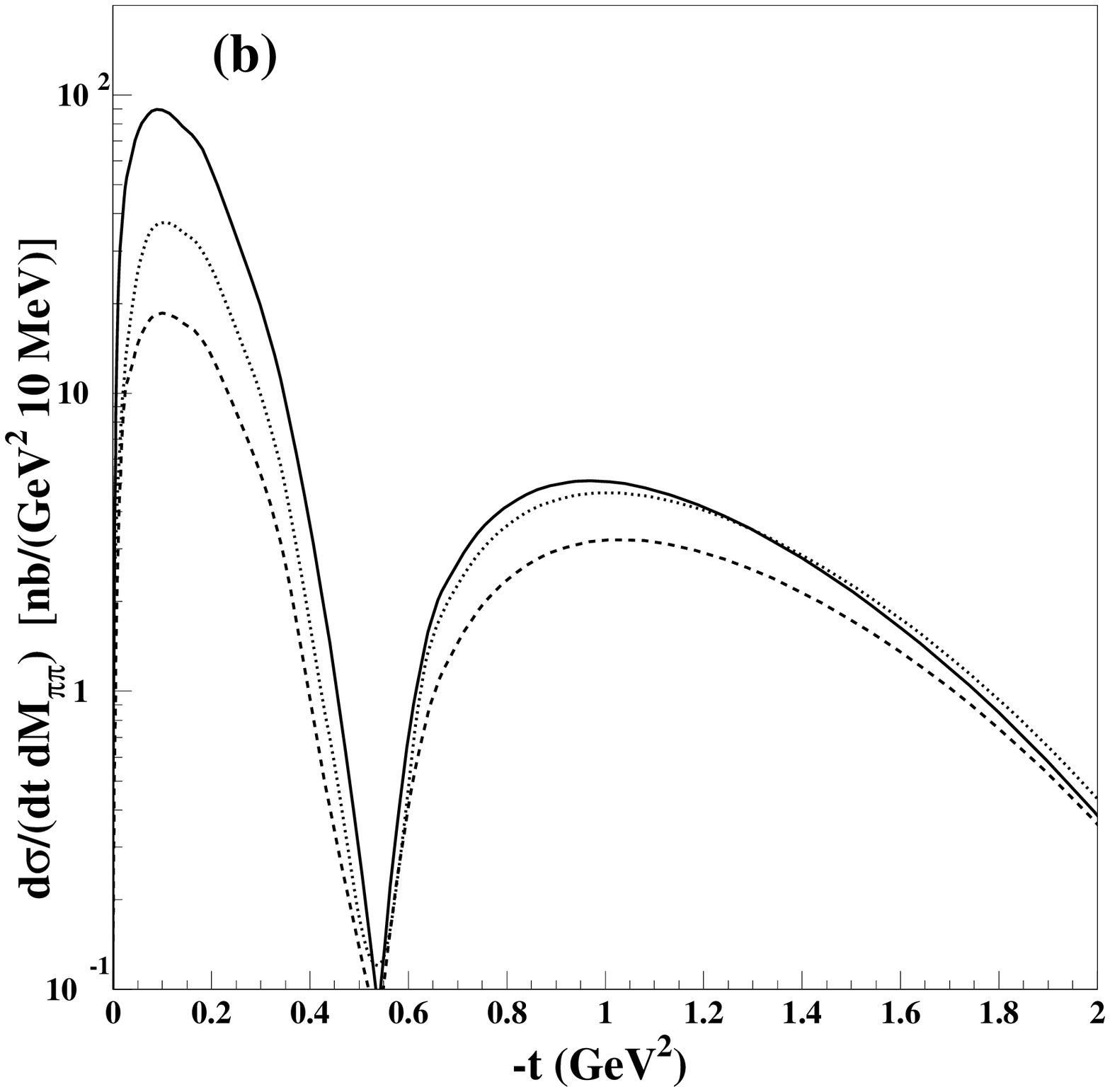}{14.5cm}{10}{145}{530}{655}{14.5cm} 

\label{fig3b}
\end{figure}

\renewcommand{\thefigure}{4(a)}
\begin{figure}
\caption{$S$-wave $\pi^{+} \pi^{-}$ invariant mass distribution at 
$E_{\gamma}^{lab} = 4.0$ GeV and $t = -0.2 \;$ GeV$^2$ showing 
sensitivity to the cut-off $\Lambda_{cut}$ using
the form factor given by Eq. (31). 
The solid, dashed, and dotted lines were calculated
with $\Lambda_{cut}$ = 0.5, 1.0, and 1.5 GeV, respectively.}

\vspace{0.5cm}

\xslide{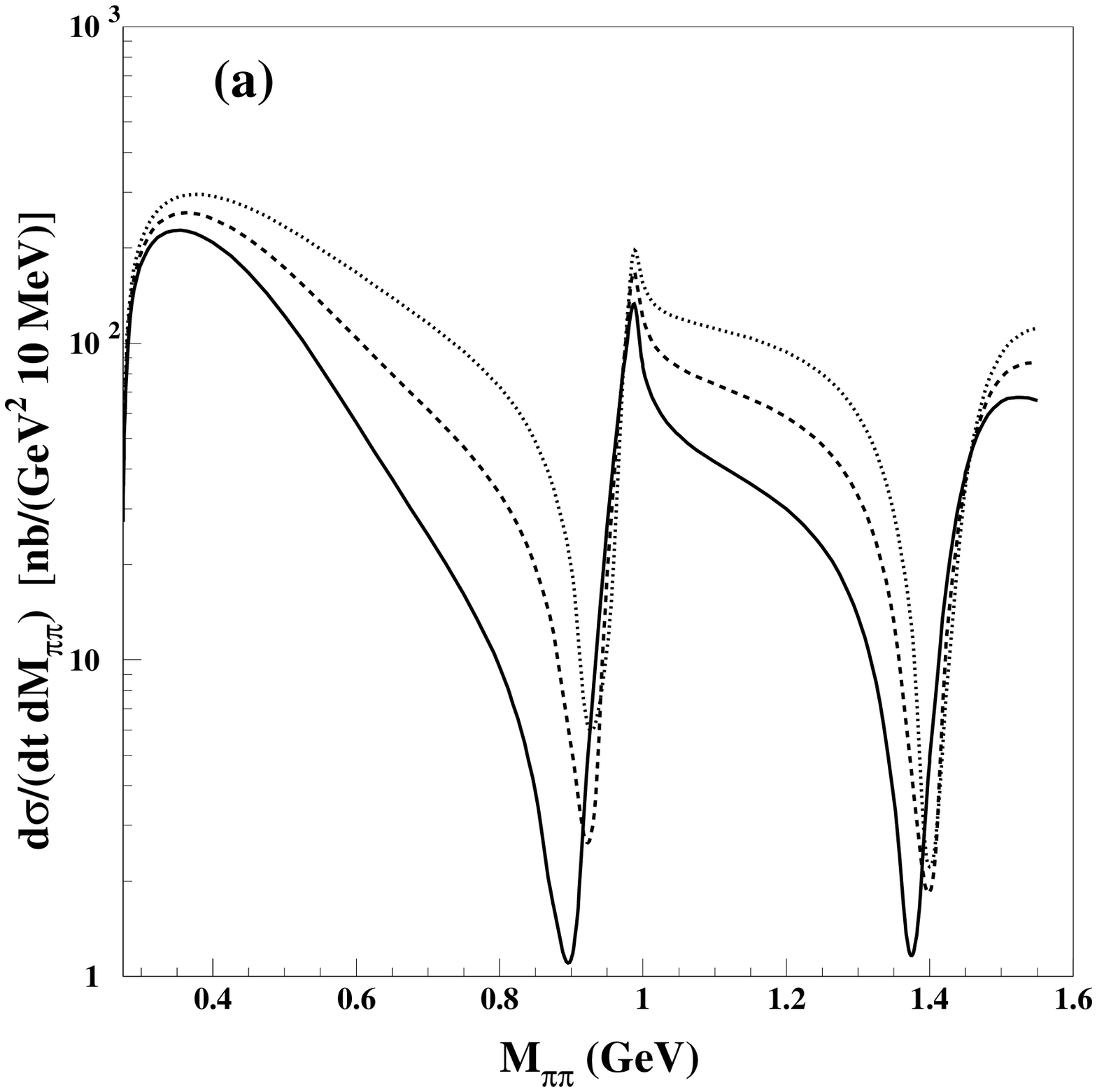}{14.5cm}{10}{145}{530}{655}{14.5cm} 

\label{fig4a}
\end{figure}

\renewcommand{\thefigure}{4(b)}
\begin{figure}
\caption{$S$-wave $\pi^{+} \pi^{-}$ invariant mass distribution at 
$E_{\gamma}^{lab} = 4.0$ GeV and $t = -0.2 \;$ GeV$^2$ showing 
sensitivity to the cut-off $\Lambda_{cut}$ using
the form factor given by Eq.(32).
The solid, dashed, and dotted lines were calculated
with $\Lambda_{cut}$ = 0.5, 1.0, and 1.5 GeV, respectively.}

\vspace{0.5cm}

\xslide{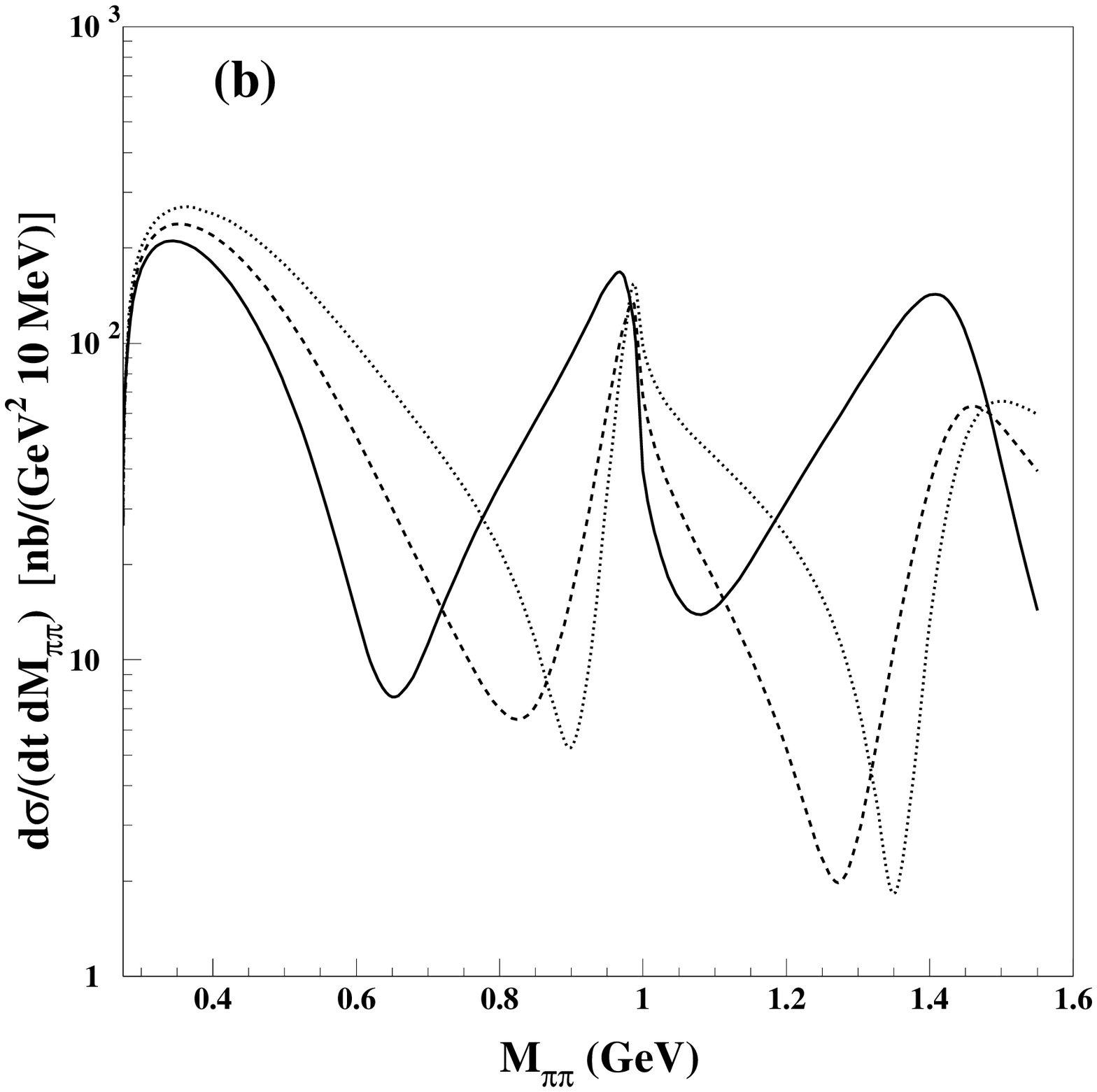}{14.5cm}{10}{145}{530}{655}{14.5cm} 

\label{fig4b}
\end{figure}
  
\renewcommand{\thefigure}{5(a)}
\begin{figure}
\caption{$S$-wave $K^{+} K^{-}$ invariant mass distribution at 
$E_{\gamma}^{lab} = 4.0 \;$ GeV and $t = -0.2 \;$ GeV$^2$. Both curves
are calculated with no final state interactions (Born cross section).  
The solid and dashed lines were calculated using Bonn and Spanish 
parameters, respectively.}

\vspace{0.5cm}

\xslide{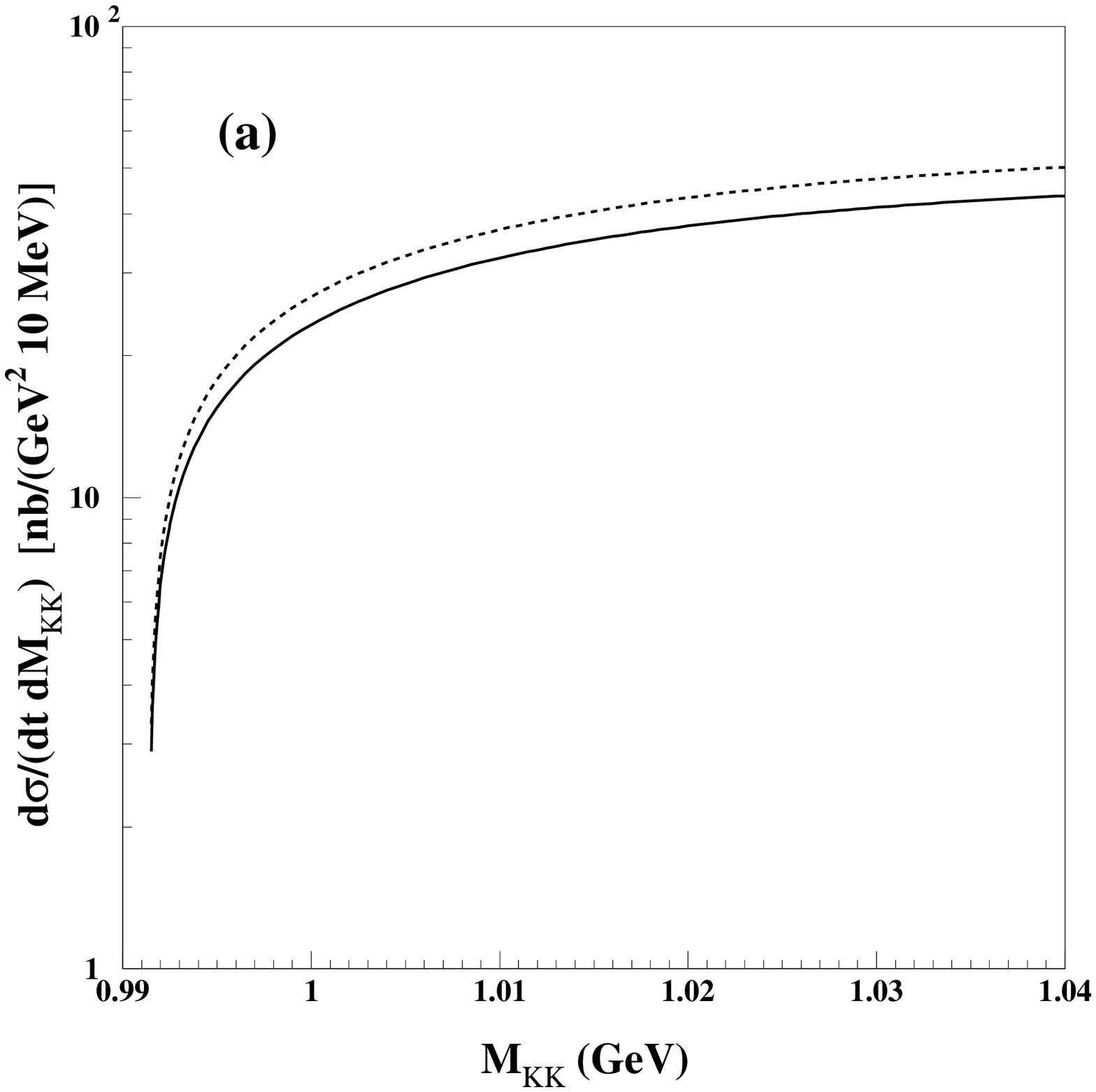}{14.5cm}{10}{145}{530}{655}{14.5cm} 

\label{fig5a}
\end{figure}

\renewcommand{\thefigure}{5(b)}
\begin{figure}
\caption{$S$-wave $K^{+} K^{-}$ invariant mass distribution at 
$E_{\gamma}^{lab} = 4.0 \;$ GeV and $t = -0.2 \;$ GeV$^2$ showing 
purely on shell FSI terms
relative to the Born cross section (dashed line). The solid line 
shows the FSI result with both $\pi \pi$ and $K\bar{K}$ 
intermediate channels whereas the dotted line represents the result with
no $\pi \pi$ coupling.}

\vspace{0.5cm}

\xslide{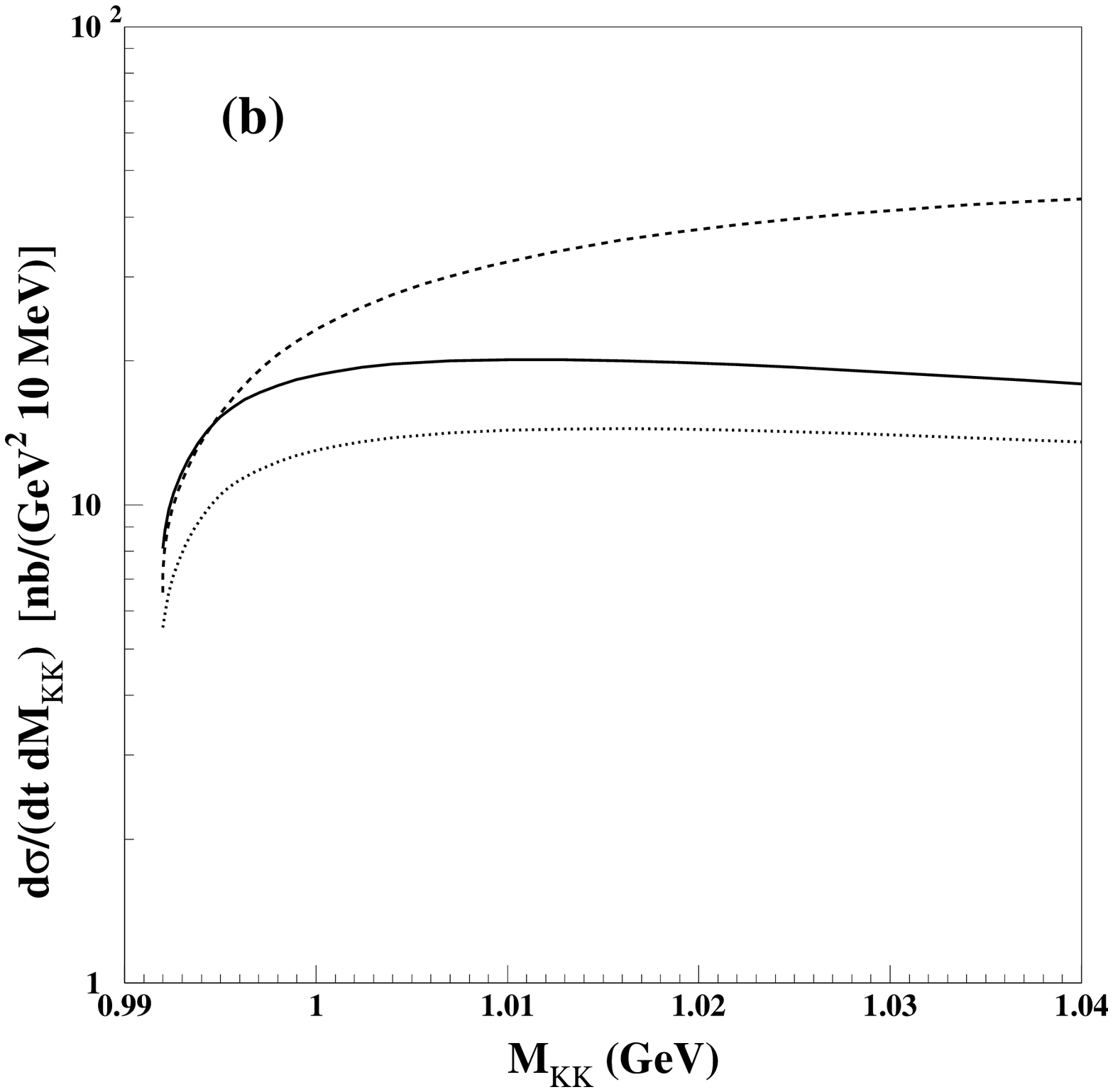}{14.5cm}{10}{145}{530}{655}{14.5cm} 

\label{fig5b}
\end{figure}
\renewcommand{\thefigure}{5(c)}
\begin{figure}
\caption{Expanded $M_{K K}$ range of 
Fig. 5(b) with the same curve labeling. }

\vspace{0.5cm}

\xslide{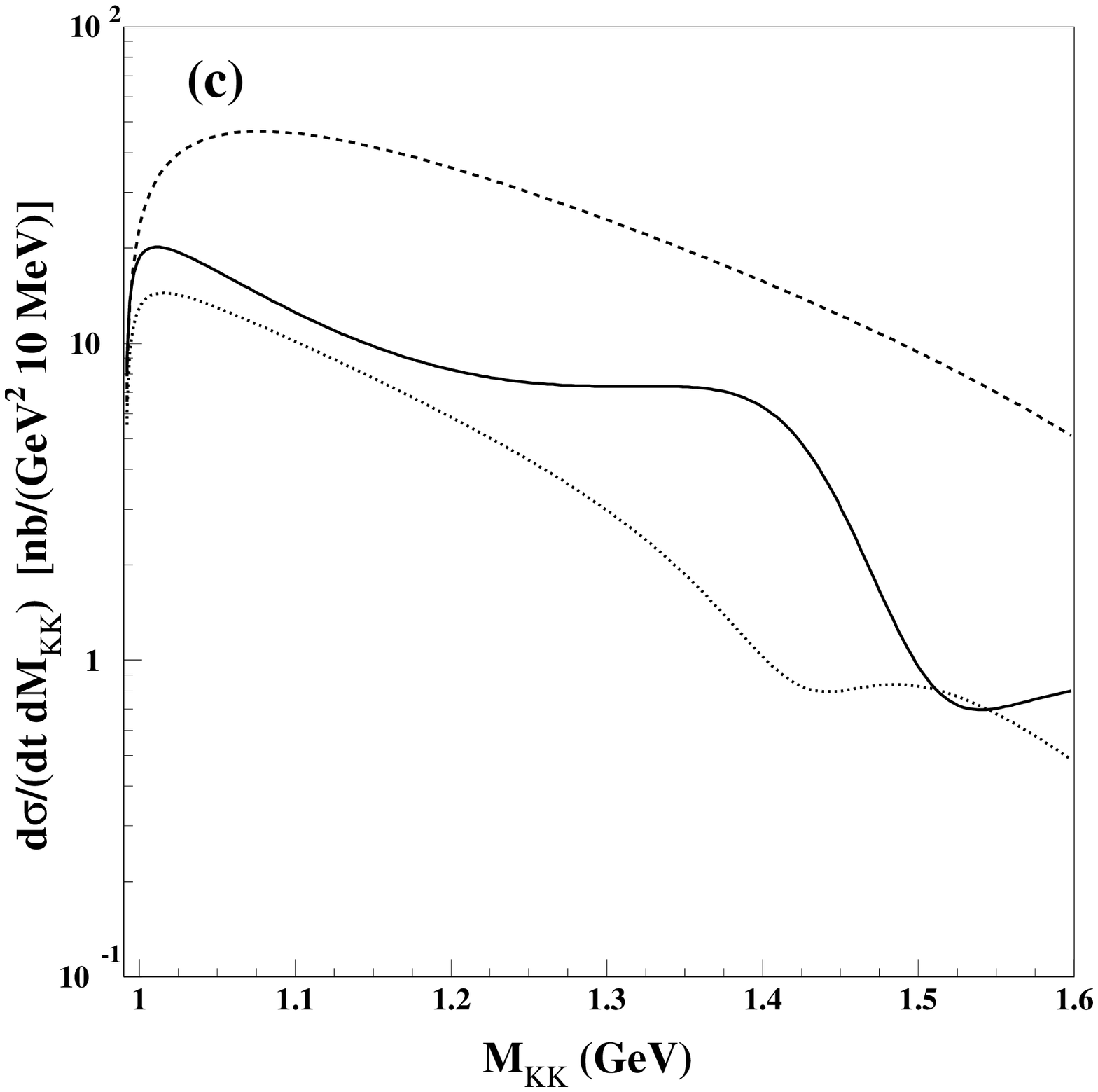}{14.5cm}{10}{145}{530}{655}{14.5cm} 

\label{fig5c}
\end{figure}

\renewcommand{\thefigure}{5(d)}
\begin{figure}
\caption{$S$-wave $K^{+} K^{-}$ invariant mass distribution at 
$E_{\gamma}^{lab} = 4.0 \;$ GeV and $t = -0.2 \;$ GeV$^2$ showing 
the full FSI terms (on--shell and off--shell)
relative to the Born cross section (dashed line). The solid line 
shows the full FSI result with both $\pi \pi$ and $K\bar{K}$ 
intermediate channels whereas the dotted line represents the result with
no $\pi \pi$ coupling.}

\vspace{0.5cm}

\xslide{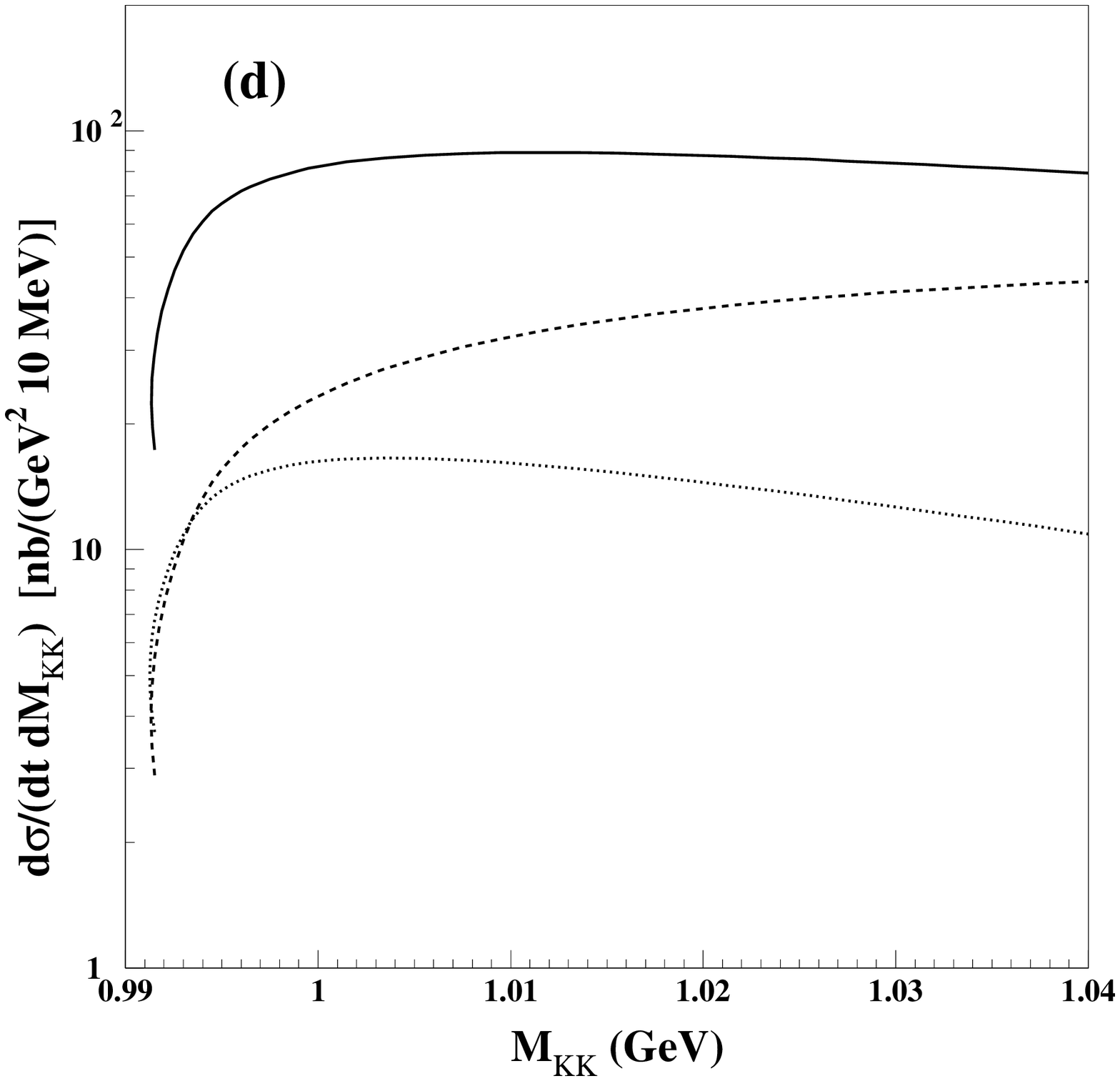}{14.5cm}{10}{145}{530}{655}{14.5cm} 

\label{fig5d}
\end{figure}

\renewcommand{\thefigure}{5(e)}
\begin{figure}
\caption{Expanded $M_{K K}$ range of 
Fig. 5(d) with the same curve labeling. }

\vspace{0.5cm}

\xslide{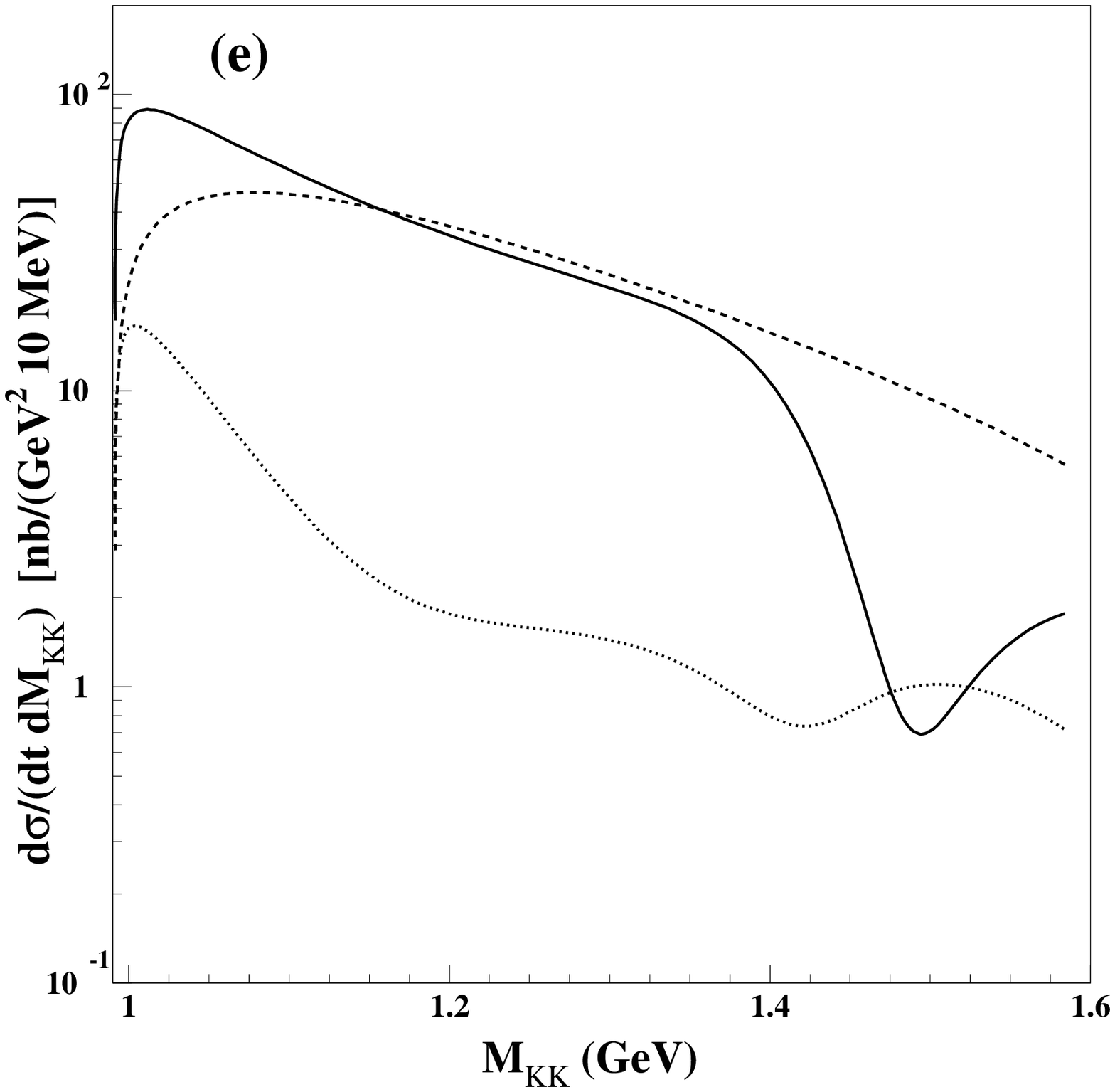}{14.5cm}{10}{145}{530}{655}{14.5cm} 

\label{fig5e}
\end{figure}

\renewcommand{\thefigure}{6}
\begin{figure}
\caption{t-dependence of the $S$-wave $K^{+} K^{-}$ photoproduction 
cross section at $E_{\gamma}^{lab} = 4.0 \;$ GeV with
$M_{cut} = 1.04 \;$ GeV.
Solid and dashed lines represent the cross sections with
full FSI terms 
using normal and Regge propagators, respectively.
Dotted and dotted-dashed lines represent the Born results
using normal and Regge propagators, respectively.
Experimental data are taken from Ref. [1].}

\vspace{0.5cm}

\xslide{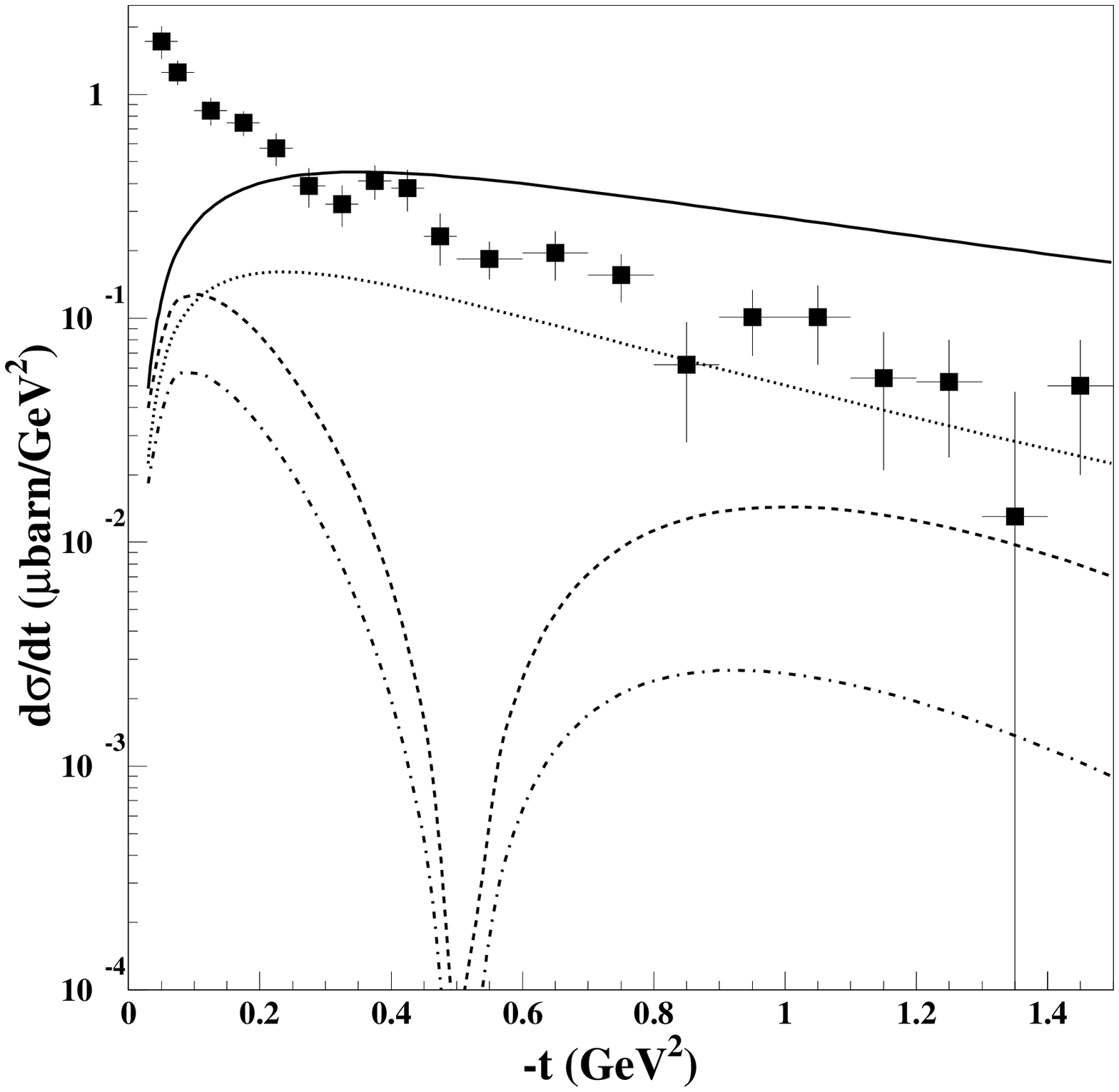}{14.5cm}{10}{145}{530}{655}{14.5cm} 

\label{fig6}
\end{figure}

\renewcommand{\thefigure}{7(a)}
\begin{figure}
\caption{$M_{K K}$ dependence of the $S$-wave $K^{+} K^{-}$ 
photoproduction cross section 
at $E_{\gamma}^{lab} = 4.0$ GeV with $t_{cut} = -1.5 \;$ GeV$^2$.
Solid and dashed lines represent the cross sections with full FSI terms
using normal and Regge propagators, respectively.
Dotted and dotted-dashed lines represent the Born results
using normal and Regge propagators, respectively.}

\vspace{0.5cm}

\xslide{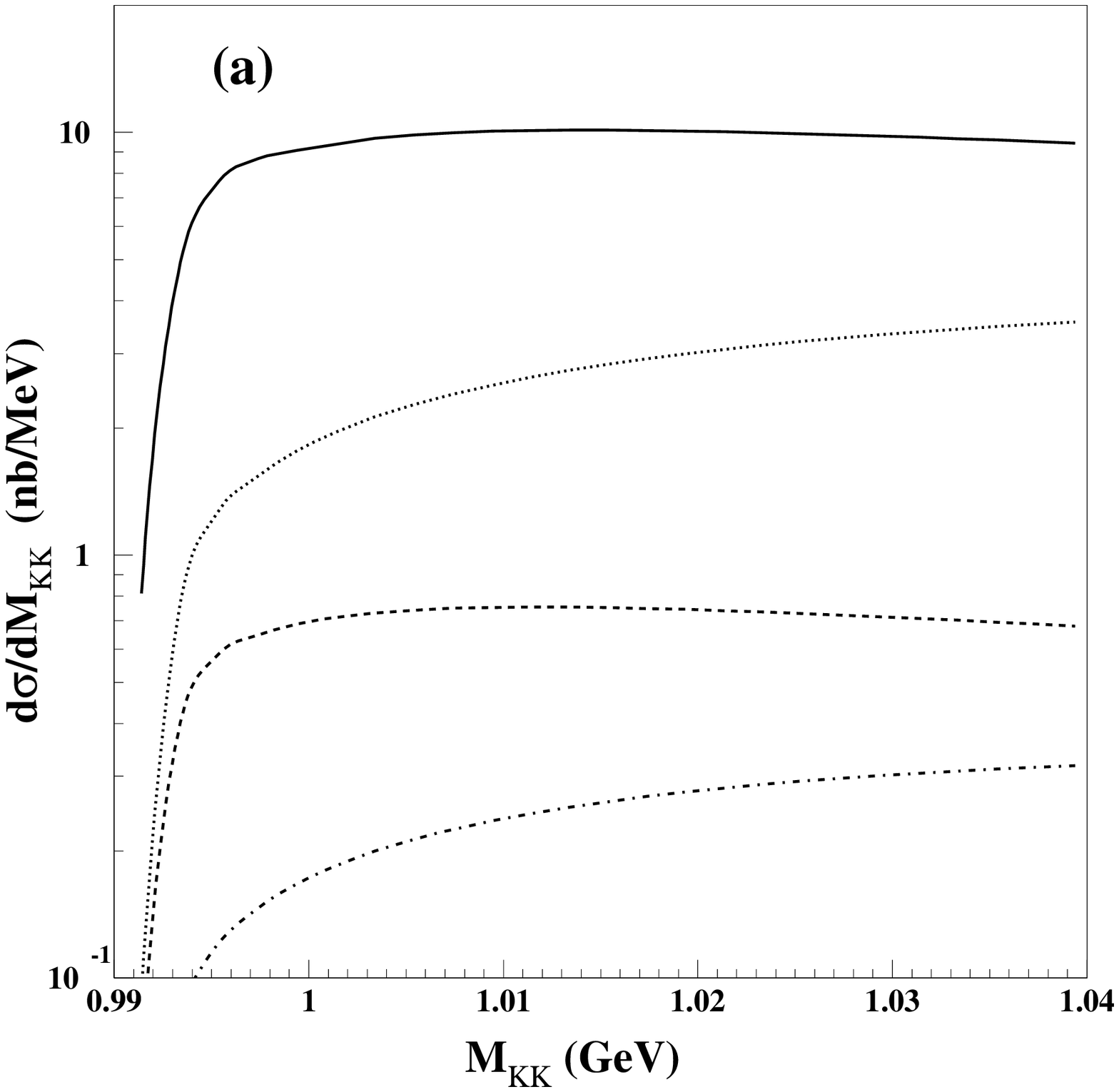}{14.5cm}{10}{145}{530}{655}{14.5cm} 

\label{fig7a}
\end{figure}

\renewcommand{\thefigure}{7(b)}
\begin{figure}
\caption{$M_{K K}$ dependence of $S$-wave $K^{+} K^{-}$ photoproduction cross section 
at $E_{\gamma}^{lab} = 4.0$ GeV
with $t_{cut} = -0.2 \;$ GeV$^2$ and the same curve labeling as in Fig. 7(a).}

\vspace{0.5cm}

\xslide{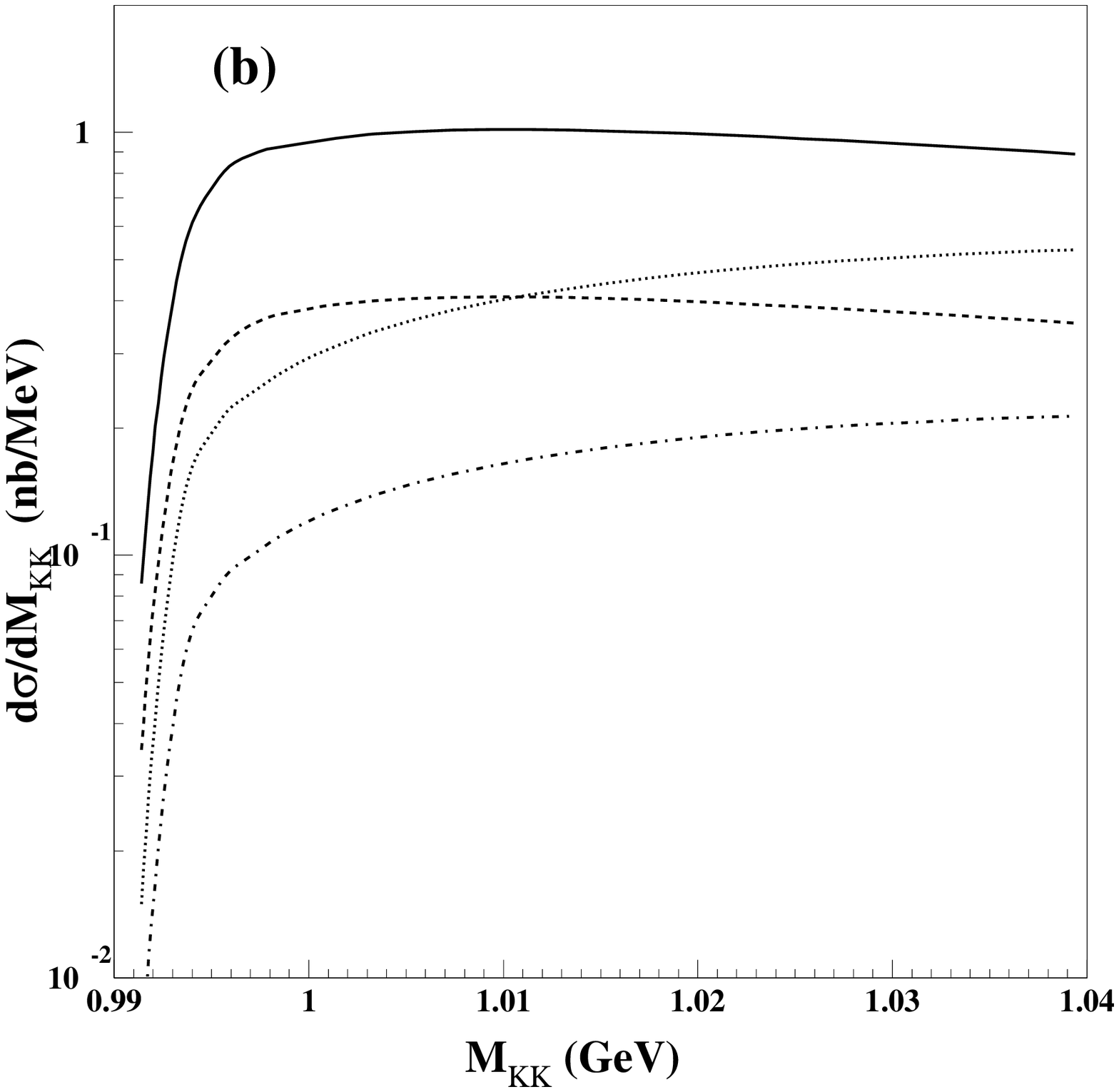}{14.5cm}{10}{145}{530}{655}{14.5cm} 

\label{fig7b}
\end{figure}

\renewcommand{\thefigure}{8(a)}
\begin{figure}
\caption{Photon energy dependence of the \kk photoproduction cross section at 
$M_{KK} =1$ GeV and $t=-0.2$ GeV$^2$ (solid line) and $t=-1$ GeV$^2$ 
(dashed line). Normal $\rho$, $\omega$ propagators are used.}

\vspace{0.5cm}

\xslide{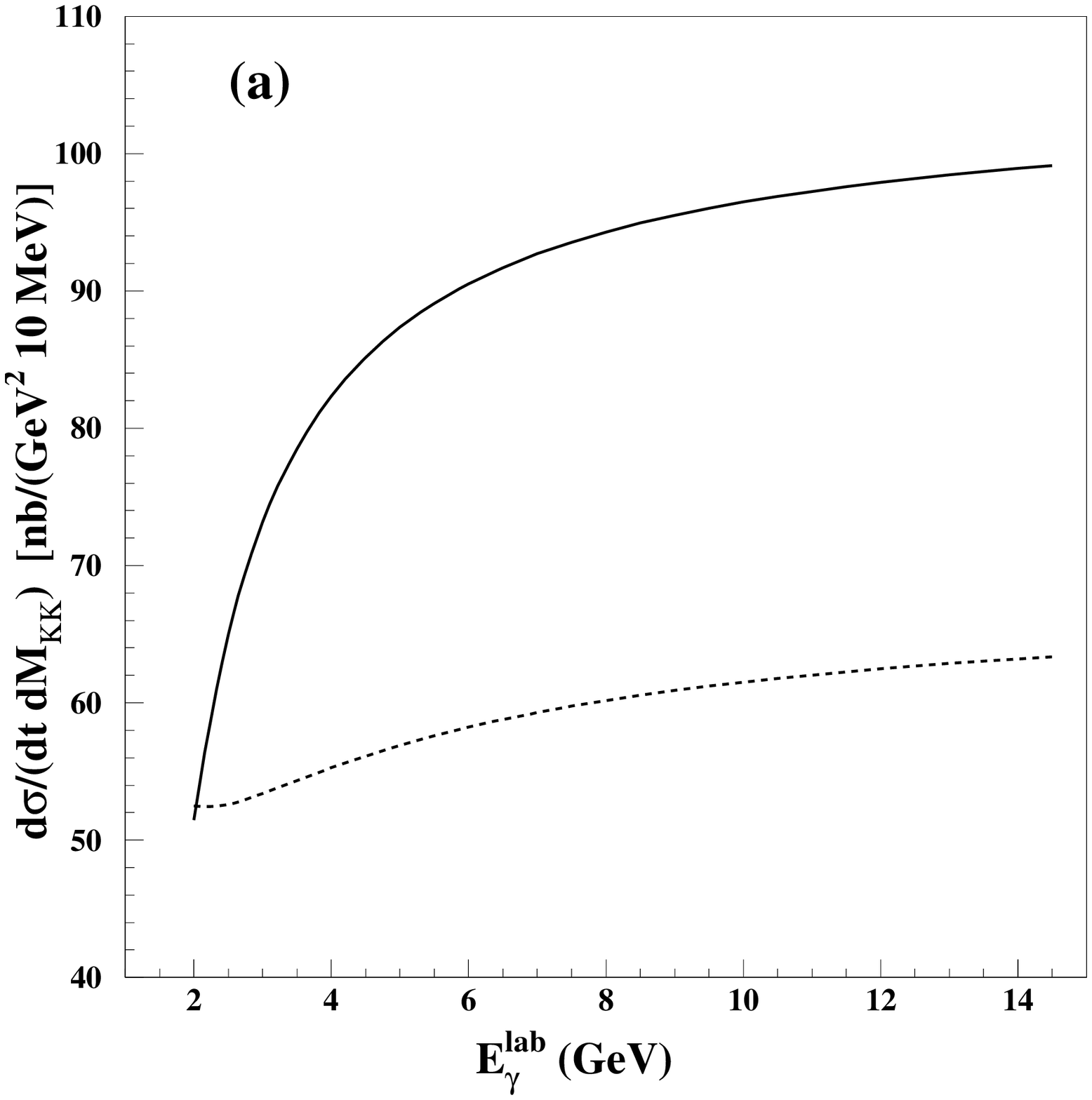}{14.5cm}{10}{145}{530}{655}{14.5cm} 

\label{fig8a}
\end{figure}

\renewcommand{\thefigure}{8(b)}
\begin{figure}
\caption{Same as in Fig. 8(a) but Regge $\rho$, $\omega$ propagators
were used.}

\vspace{0.5cm}

\xslide{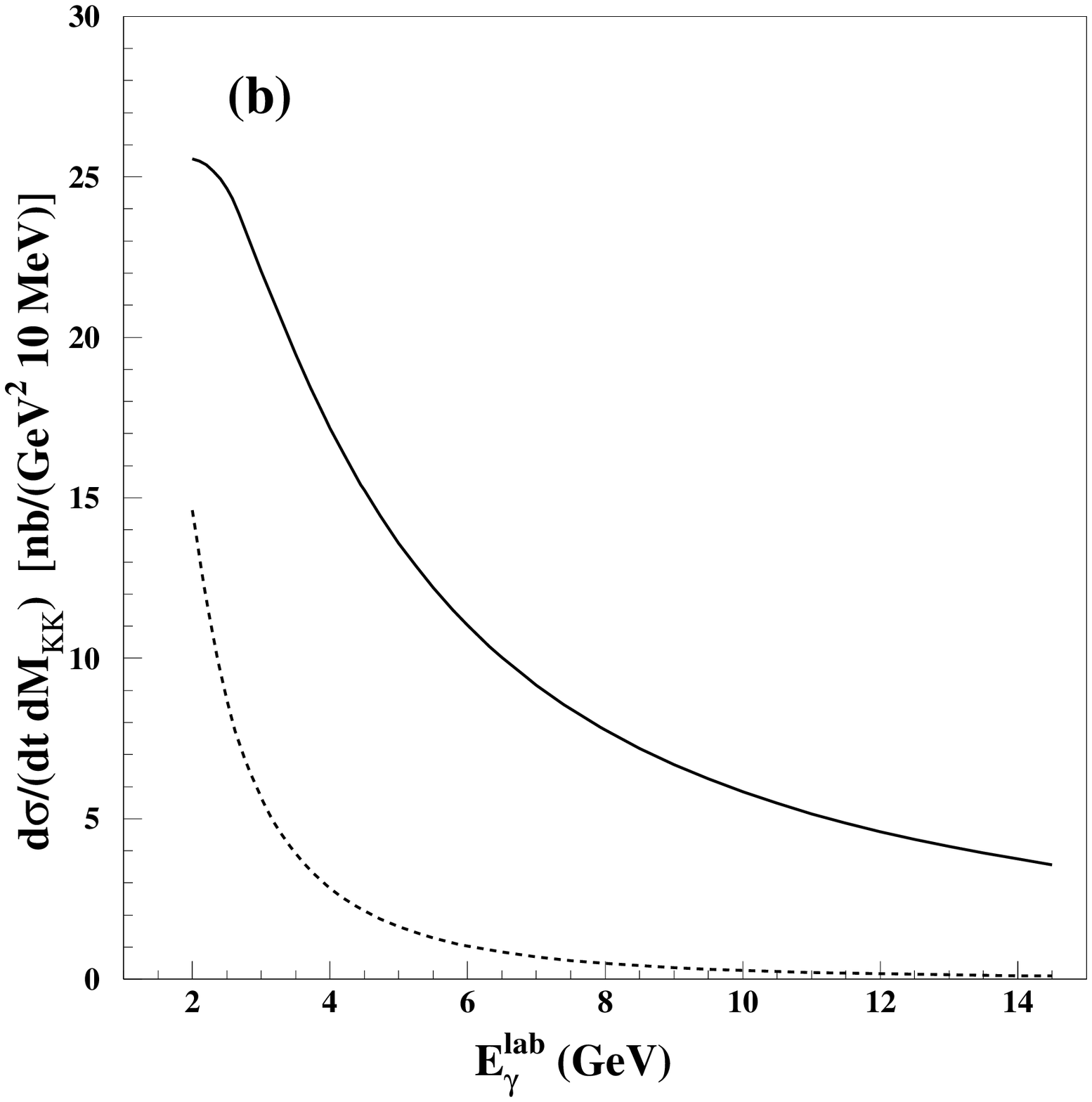}{14.5cm}{10}{145}{530}{655}{14.5cm} 

\label{fig8b}
\end{figure}

\end{document}